Variational Calculation for the Equation of State of Nuclear Matter at Finite Temperatures

H. Kanzawa[a], K. Oyamatsu[b], K. Sumiyoshi[c, d] and M. Takano [e]

a Department of Physics, Science and Engineering, Waseda University, 3-4-1 Okubo Shinjuku-ku, Tokyo 169-8555, Japan
b Department of Media Production and Theories, Aichi Shukutoku University, Nagakute-cho, Aichi 480-1197, Japan
c Numazu College of Technology, Ooka 3600, Numazu, Shizuoka 410-8501, Japan
d Division of Theoretical Astronomy, National Astronomical Observatory of Japan, 2-21-1, Osawa, Mitaka, Tokyo 181-8588, Japan
e Advanced Research Institute for Science and Engineering, Waseda University, 3-4-1 Okubo Shinjuku-ku, Tokyo 169-8555, Japan

An equation of state (EOS) for uniform nuclear matter is constructed at zero and finite temperatures with the variational method starting from the realistic nuclear Hamiltonian composed of the Argonne V18 and UIX potentials. The energy is evaluated in the two-body cluster approximation with the three-body-force contribution treated phenomenologically so as to reproduce the empirical saturation conditions. The obtained energies for symmetric nuclear matter and neutron matter at zero temperature are in fair agreement with those by Akmal, Pandharipande and Ravenhall, and the maximum mass of the neutron star is 2.2 $M_{\odot}$. At finite temperatures, a variational method by Schmidt and Pandharipande is employed to evaluate the free energy, which is used to derive various thermodynamic quantities of nuclear matter necessary for supernova simulations. The result of this variational method at finite temperatures is found to be self-consistent.

## 1. Introduction

The variational method [1] is one of the most powerful many-body theories, and many variational calculations have been reported for the nuclear equation of state (EOS), especially for nuclear matter at zero temperature. A typical example is by Akmal, Pandharipande and Ravenhall (APR) [2]. Starting from the modern nuclear Hamiltonian composed of the Argonne V18 (AV18) [3] two-body nuclear interaction and the Urbana IX (UIX) [4] three-nucleon interaction (TNI), APR evaluated the ground-state energies for symmetric nuclear matter and neutron matter using the Fermi hypernetted chain (FHNC) method [5].

On the other hand, there are rather few examples of variational calculations for nuclear matter at finite temperatures; a familiar example is by Friedman and Pandharipande (FP) [6] in 1981. Based on the Urbana V14 two-body nuclear interaction with phenomenological corrections for TNI, FP evaluated the free energies using the prescription proposed by Schmidt and Pandharipande (SP) [7]. Another example is the lowest-order constrained variational calculations for asymmetric nuclear matter at finite temperatures [8]. In their calculations, the empirical saturation point is not reproduced because TNI is not taken into account.

Meanwhile, the necessity of researches on nuclear matter for astrophysics is increasing. Recent progress of supernova simulations revives the importance of the nuclear EOS at finite temperatures in the mechanisms of supernova explosions [9]. The EOS is essential to determine the core bounce, the propagation of shock wave, the formation of proto-neutron stars through the behavior of pressure, entropy, chemical potentials and compositions. The set of thermodynamic quantities are necessary to perform numerical simulations of hydrodynamics and neutrino transfer with the weak reaction rates.

Actually, however, the set of EOS for supernova simulations was limited. Recently, a new set of EOS has become available (SH-EOS[10]) in addition to the conventional one (LS-EOS[11]). These sets are obtained in phenomenological frameworks: an extension of the compressible liquid drop model (LS-EOS) and the relativistic mean field theory (SH-EOS). These phenomenological frameworks are adopted because one has to evaluate the quantities of nuclear matter in a wide range of density, proton mixture and temperature. In order to construct the EOS tables, the calculations are based on bulk properties with effective nuclear forces and are not directly derived from the bare nuclear forces. Since the EOS's available for supernovae are limited to these phenomenological calculations, more microscopic approaches are desirable to study the direct connection between the nuclear force and the nuclear EOS in supernovae.

Taking into account the above-mentioned situation, we undertake to construct the nuclear EOS for supernovae using a microscopic many-body approach. As the first step of this undertaking, we construct, in this paper, an EOS for uniform nuclear matter at finite temperatures using the variational method, starting from the modern nuclear Hamiltonian AV18+UIX. In order to treat the nuclear EOS at zero temperature and at finite temperatures in a consistent way, we employ a relatively simple variational method. We develop the two-body cluster approximation with appropriately chosen subsidiary conditions instead of directly treating higher-order cluster terms. We require that the nuclear EOS at zero temperature must reproduce empirical saturation condition. The saturation density $\rho_0$, saturation energy $E_0/N$, incompressibility $K$ and symmetry energy $E_s/N$ must be consistent with the empirical values. We also treat the expectation value of the UIX potential somewhat phenomenologically so that the total energy per nucleon $E/N$ reproduces the empirical values at the saturation point. In the case of nuclear matter at finite temperatures, we follow the procedure by SP. We study the thermodynamic quantities such as pressure, and examine the self-consistency of the results obtained by SP.

In Section 2, we calculate $E/N$ for symmetric nuclear matter and neutron matter at zero temperature in two steps: First, we evaluate the energies without TNI in the two-body cluster approximation, and then we take into account TNI phenomenologically. In Section 3, we perform the variational calculation for nuclear matter at finite temperatures following the prescription by SP, and obtain the free energy, pressure, internal energy, entropy, effective mass and thermal energy for symmetric nuclear matter and neutron matter. Concluding remarks are given in Section 4.

## 2. Nuclear matter at zero temperature

### 2.1. Energies for nuclear matter without the three-nucleon interaction

In this section, we mainly consider symmetric nuclear matter at zero temperature; the energy for neutron matter is calculated in a similar way. We start from the Hamiltonian without the three-nucleon interaction (TNI):

$$H_2 = -\frac{\hbar^2}{2m}\sum_{i=1}^{N}\nabla_i^2 + \sum_{i<j}^{N} V_{ij}\,, \tag{1}$$

where $N$ is the total number of nucleons and $m$ is the nucleon mass. The two-body nuclear potential $V_{ij}$ is chosen as the isoscalar part of the AV18 potential

$$V_{ij} = \sum_{t=0}^{1}\sum_{s=0}^{1}\Big[V_{\mathrm{C}ts}(r_{ij}) + sV_{\mathrm{T}t}(r_{ij})S_{\mathrm{T}ij} + sV_{\mathrm{SO}t}(r_{ij})\big(\boldsymbol{L}_{ij}\cdot\mathbf{s}\big)$$
$$+V_{\mathrm{qL}ts}(r_{ij})\left|\boldsymbol{L}_{ij}\right|^2 + sV_{\mathrm{qSO}t}(r_{ij})\big(\boldsymbol{L}_{ij}\cdot\mathbf{s}\big)^2\Big]P_{tsij}\,. \tag{2}$$

Here, $S_{\mathrm{T}ij}$ is the tensor operator, and $\boldsymbol{L}_{ij}$ is the relative orbital angular momentum operator of the ($i$, $j$) nucleon pair. The isospin-spin projection operator $P_{tsij}$ projects the ($i$, $j$) nucleon pair states onto the triplet-odd (($t$, $s$)=(1, 1)), singlet-even (($t$, $s$)=(1, 0)), triplet-even (($t$, $s$)=(0, 1)) or singlet-odd (($t$, $s$)=(0, 0)) states. The remaining isovector and isotensor terms in the AV18 are omitted as in the FHNC calculations by APR, because their contribution to the energy is expected to be small.

For the variational wave function $\Psi$, we assume the Jastrow-type function which is the same as used by APR:

$$\psi = \mathrm{Sym}\left[\prod_{i<j} f_{ij}\right]\Phi_{\mathrm{F}}\,, \tag{3}$$

where $\Phi_{\mathrm{F}}$ is the Fermi-gas wave function at zero temperature and Sym[ ] is the symmetrizer with respect to the order of the factors in the products. The two-body correlation function $f_{ij}$ is taken as

$$f_{ij} = \sum_{t=0}^{1}\sum_{s=0}^{1}\Big[f_{\mathrm{C}ts}(r_{ij}) + sf_{\mathrm{T}t}(r_{ij})S_{\mathrm{T}ij} + sf_{\mathrm{SO}t}(r_{ij})\big(\boldsymbol{L}_{ij}\cdot\boldsymbol{s}\big)\Big]P_{tsij}\,. \tag{4}$$

It is noted that the central, tensor and spin-orbit correlation functions $f_{\mathrm{C}ts}(r)$, $f_{\mathrm{T}t}(r)$ and $f_{\mathrm{SO}t}(r)$ in Eq. (4) are treated as variational functions.

Using the wave function Eq. (3), the expectation value of $H_2$, i.e., $\langle H_2\rangle = \langle\Psi|H_2|\Psi\rangle/\langle\Psi|\Psi\rangle$ is cluster-expanded into a series of many-body cluster terms. In this study, we employ the two-body cluster approximation, i.e., we truncate the series and keep only the one-body and two-body cluster terms. Then, $E_2/N$, which is $\langle H_2\rangle/N$ in the two-body cluster approximation, is expressed as

$$\frac{E_2}{N} = \frac{3}{5}E_{\mathrm{F}} + 2\pi\rho\sum_{t=0}^{1}\sum_{s=0}^{1}\int_0^\infty r^2 dr\big[F_{2ts}(r)V_{\mathrm{C}ts}(r) + sF_{2\mathrm{T}t}(r)V_{\mathrm{T}t}(r)$$
$$+ sF_{2\mathrm{SO}t}(r)V_{\mathrm{SO}t}(r) + F_{2\mathrm{qL}ts}(r)V_{\mathrm{qL}ts}(r) + sF_{2\mathrm{qSO}t}(r)V_{\mathrm{qSO}t}(r)\big]$$
$$+ 2\pi\rho\frac{\hbar^2}{m}\sum_{t=0}^{1}\sum_{s=0}^{1}\int_0^\infty r^2 dr\left\{\left[\frac{df_{\mathrm{C}ts}(r)}{dr}\right]^2 F_{\mathrm{F}ts}(r)\right.$$
$$\left. + s\left\{8\left[\frac{df_{\mathrm{T}t}(r)}{dr}\right]^2 + \frac{48}{r^2}\big[f_{\mathrm{T}t}(r)\big]^2\right\}F_{\mathrm{F}ts}(r) + \frac{2}{3}s\left[\frac{df_{\mathrm{SO}t}(r)}{dr}\right]^2 F_{\mathrm{qF}ts}(r)\right\}. \tag{5}$$

The first term on the right-hand side of Eq. (5) is the one-body kinetic energy, and $E_{\mathrm{F}} = \hbar^2 k_{\mathrm{F}}^2/(2m)$ is the Fermi energy with $k_{\mathrm{F}}$ being the Fermi wave number; in the case of symmetric nuclear matter, $k_{\mathrm{F}} = (3\pi^2\rho/2)^{1/3}$ with the number density $\rho$. The second term in Eq. (5) represents the potential energy, in which $F_{2ts}(r)$, $F_{2\mathrm{T}t}(r)$, $F_{2\mathrm{SO}t}(r)$, $F_{2\mathrm{qL}ts}(r)$ and $F_{2\mathrm{qSO}t}(r)$ are expressed as

$$F_{2ts}(r) = \Big\{\big[f_{\mathrm{C}ts}(r)\big]^2 + 8s\big[f_{\mathrm{T}t}(r)\big]^2\Big\}F_{\mathrm{F}ts}(r) + \frac{2}{3}s\big[f_{\mathrm{SO}t}(r)\big]^2 F_{\mathrm{qF}ts}(r)\,, \tag{6}$$

$$F_{2\mathrm{T}t}(r) = 16 f_{\mathrm{T}t}(r)\big[f_{\mathrm{C}t1}(r) - f_{\mathrm{T}t}(r)\big]F_{\mathrm{F}t1}(r) - \frac{2}{3}\big[f_{\mathrm{SO}t}(r)\big]^2 F_{\mathrm{qF}t1}(r)\,, \tag{7}$$

$$F_{2\mathrm{SO}t}(r)=\left\{\frac{4}{3}f_{\mathrm{SO}t}(r)\left[f_{\mathrm{C}t1}(r)-f_{\mathrm{T}t}(r)\right]-\frac{1}{3}\left[f_{\mathrm{SO}t}(r)\right]^2\right\}F_{\mathrm{qF}t1}(r)$$

$$-24\left[f_{\mathrm{T}t}(r)\right]^2F_{\mathrm{F}t1}(r), \tag{8}$$

$$F_{2\mathrm{qL}ts}(r)=48s\left[f_{\mathrm{T}t}(r)\right]^2F_{\mathrm{F}ts}(r)+\left\{\left[f_{\mathrm{C}ts}(r)\right]^2+8s\left[f_{\mathrm{T}t}(r)\right]^2\right\}F_{\mathrm{qF}ts}(r)$$

$$-\frac{2}{3}s\left[f_{\mathrm{SO}t}(r)\right]^2F_{\mathrm{bF}ts}(r), \tag{9}$$

$$F_{2\mathrm{qSO}t}(r)=72\left[f_{\mathrm{T}t}(r)\right]^2F_{\mathrm{F}t1}(r)+\left\{\frac{2}{3}\left[f_{\mathrm{C}t1}(r)\right]^2-\frac{4}{3}f_{\mathrm{C}t1}(r)f_{\mathrm{T}t}(r)-\frac{2}{3}f_{\mathrm{C}t1}(r)f_{\mathrm{SO}t}(r)\right.$$

$$\left.+\frac{20}{3}\left[f_{\mathrm{T}t}(r)\right]^2+\frac{8}{3}f_{\mathrm{T}t}(r)f_{\mathrm{SO}t}(r)\right\}F_{\mathrm{qF}t1}(r)+\frac{20}{3}\left[f_{\mathrm{SO}t}(r)\right]^2F_{\mathrm{bF}t1}(r), \tag{10}$$

where $F_{\mathrm{F}ts}(r)$, $F_{\mathrm{qF}ts}(r)$ and $F_{\mathrm{bF}ts}(r)$ are

$$F_{\mathrm{F}ts}(r_{12})=\Omega^2\sum_{\text{isospin}}\sum_{\text{spin}}\int \Phi_{\mathrm{F}}^{\dagger}P_{ts12}\Phi_{\mathrm{F}}d\boldsymbol{r}_3d\boldsymbol{r}_4\cdots d\boldsymbol{r}_N$$

$$=\frac{(2t+1)(2s+1)}{16}\left\{1-(-1)^{t+s}\left[3\frac{j_1(k_{\mathrm{F}}r)}{k_{\mathrm{F}}r}\right]^2\right\}, \tag{11}$$

$$F_{\mathrm{qF}ts}(r_{12})=\Omega^2\sum_{\text{isospin}}\sum_{\text{spin}}\int \Phi_{\mathrm{F}}^{\dagger}\left|\boldsymbol{L}_{12}\right|^2P_{ts12}\Phi_{\mathrm{F}}d\boldsymbol{r}_3d\boldsymbol{r}_4\cdots d\boldsymbol{r}_N$$

$$=\frac{(2t+1)(2s+1)}{16}\left[\frac{(k_{\mathrm{F}}r)^2}{5}+(-1)^{t+s}9j_2(k_{\mathrm{F}}r)\frac{j_1(k_{\mathrm{F}}r)}{k_{\mathrm{F}}r}\right], \tag{12}$$

$$F_{\mathrm{bF}ts}(r_{12})=\Omega^2\sum_{\text{isospin}}\sum_{\text{spin}}\int \Phi_{\mathrm{F}}^{\dagger}\left|\boldsymbol{L}_{12}\right|^4P_{ts12}\Phi_{\mathrm{F}}d\boldsymbol{r}_3d\boldsymbol{r}_4\cdots d\boldsymbol{r}_N$$

$$=\frac{(2t+1)(2s+1)}{16}\left[\frac{2}{5}(k_{\mathrm{F}}r)^2+\frac{12}{175}(k_{\mathrm{F}}r)^4\right.$$

$$\left.-(-1)^{t+s}9\left\{j_1(k_{\mathrm{F}}r)j_3(k_{\mathrm{F}}r)+\left[j_2(k_{\mathrm{F}}r)\right]^2-2\frac{j_1(k_{\mathrm{F}}r)j_2(k_{\mathrm{F}}r)}{k_{\mathrm{F}}r}\right\}\right]. \tag{13}$$

Here, $\Omega$ is the volume of the system and $\Sigma$ represents the summation over the isospin- and spin- coordinates of all the nucleons. The last term on the right-hand side of Eq. (5) represents the kinetic energy caused by the correlation between nucleons.

Since $E_2/N$ is an explicit functional of the variational functions $f_{\mathrm{C}ts}(r)$, $f_{\mathrm{T}t}(r)$ and $f_{\mathrm{SO}t}(r)$, the full minimization of $E_2/N$ can be performed by solving the Euler-Lagrange (EL) equations derived from Eq. (5). However, it is known that the minimized energies without any constraints become extremely large negative values[12] because of the truncation of the three-body and higher-order cluster terms; hereafter we call them many-body cluster terms. In order to avoid such an unreasonable result, we impose two conditions on the variational functions as explained in the following.

The first condition is the extended Mayer's condition expressed as

$$4\pi\rho\int_0^\infty [F_{ts}(r)-F_{\mathrm{F}ts}(r)]r^2dr=0 \qquad (t,s=0,1), \tag{14}$$

where $F_{ts}(r)$ is the isospin-spin-dependent radial distribution function,

$$F_{ts}(r_{12})=\Omega^2\sum_{\text{isospin}}\sum_{\text{spin}}\int \Psi^\dagger P_{ts12}\Psi d\boldsymbol{r}_3 d\boldsymbol{r}_4\cdots d\boldsymbol{r}_N\,. \tag{15}$$

It is noted that the condition (14) satisfies the original Mayer's condition [13],

$$1+4\pi\rho\int_0^\infty [F(r)-1]r^2dr=0\,, \tag{16}$$

where $F(r)$ is the sum of the four isospin-spin-dependent radial distribution functions $F_{ts}(r)$ ($t$, $s$ = 0, 1). In the two-body cluster approximation, $F_{ts}(r)$ reduces to $F_{2ts}(r)$ shown in Eq. (6), and, correspondingly, the extended Mayer's condition in this approximation is expressed as Eq. (14) with $F_{ts}(r)$ replaced by $F_{2ts}(r)$.

The second condition is the healing-distance condition, where the healing distance $r_\mathrm{h}$ is defined as

$$f_{\mathrm{C}ts}(r)=1, \quad f_{\mathrm{T}t}(r)=0, \quad f_{\mathrm{SO}t}(r)=0 \qquad (r\ge r_\mathrm{h}). \tag{17}$$

Namely, the correlation between two nucleons vanishes at the distance between two nucleons $r$ larger than $r_h$. Here, $r_h$ is assumed to be common to $f_{Cts}(r)$, $f_{Tt}(r)$ and $f_{SOt}(r)$. In addition, we assume that $r_h$ is proportional to the mean distance between two nucleons in nuclear matter, i.e.,

$$r_h = a_h r_0, \tag{18}$$

where $a_h$ is a coefficient and $r_0$ is the radius of the sphere of volume $1/\rho$,

$$r_0 = \left( \frac{3}{4\pi\rho} \right)^{1/3}. \tag{19}$$

Without the healing-distance condition, the minimized energy becomes negatively large, and correspondingly, $r_h$ becomes extremely large, because of the lack of the many-body cluster terms. If the many-body cluster terms are taken into account properly, $r_h$ will remain in a reasonable range. In this study, instead of treating the many-body cluster terms directly, we impose the healing-distance condition. Namely, we regard this condition as an effect of the many-body cluster terms.

Under these two conditions, $E_2/N$ is minimized: The EL equations for $f_{Cts}(r)$, $f_{Tt}(r)$ and $f_{SOt}(r)$ are derived with the extended Mayer's condition taken into account by the Lagrange multiplier method, and the EL equations are solved with Eq. (17) as boundary conditions. In order to choose the value of $a_h$, we rely on the FHNC calculation of $\langle H_2 \rangle/N$ by APR as a typical example of the sophisticated many-body calculations. Namely, the value of $a_h$ is determined so that $E_2/N$ for symmetric nuclear matter is close to the corresponding $\langle H_2 \rangle/N$ by APR. The value of $a_h$ obtained in this way is $a_h = 1.76$.

For neutron matter, $E_2/N$ is calculated in a way similar to that in the case of symmetric nuclear matter explained above. The value of $a_h$ for neutron matter is simply assumed to be the same as for symmetric nuclear matter, i.e., $a_h = 1.76$.

In Fig. 1, $E_2/N$ for symmetric nuclear matter and neutron matter are plotted, and compared with $\langle H_2 \rangle/N$ by APR. It is seen that $E_2/N$ for symmetric nuclear matter is in good agreement with $\langle H_2 \rangle/N$ by APR, which implies that our prescription is appropriate. It is also remarkable that, even in the case of neutron matter, $E_2/N$ is close to $\langle H_2 \rangle/N$ by APR. Here, we note that, compared with $E_2/N$ in the case of neutron matter, the gradient of $\langle H_2 \rangle/N$ changes discontinuously at $\rho \sim 0.6$ fm$^{-3}$. According to APR, this behavior of $\langle H_2 \rangle/N$ is due to a transition to a $\pi^0$ condensation phase. In the FHNC calculation by APR, the healing distances for tensor correlation function and the other correlation functions are treated as independent variational parameters. Therefore, the transition to the $\pi^0$ condensation phase can be seen as the large increase in the tensor healing distance at $\rho \sim 0.6$ fm$^{-3}$. In the case of the present study, the healing distance is assumed to be the same for all the correlation functions; thus the similar behavior is not expected in $E_2/N$.

## 2.2 Energies for nuclear matter with the three-body interaction

In the next step, we consider the contribution of TNI and other corrections rather phenomenologically so that the total energy per nucleon $E/N$ reproduces the empirical saturation data. We start from the UIX potential which consists of two parts: the two-pion exchange part $V_{ijk}^{2\pi}$, and the repulsive part $V_{ijk}^{\mathrm{R}}$,

$$V_{ijk} = V_{ijk}^{2\pi} + V_{ijk}^{\mathrm{R}}. \tag{20}$$

Explicitly, they are

$$V_{ijk}^{2\pi} = A \sum_{\mathrm{cyc}} \left[ \left\{ x_{ij}, x_{ik} \right\} \left\{ \tau_i \cdot \tau_j, \tau_i \cdot \tau_k \right\} + \frac{1}{4} \left[ x_{ij}, x_{ik} \right] \left[ \tau_i \cdot \tau_j, \tau_i \cdot \tau_k \right] \right], \tag{21}$$

$$V_{ijk}^{\mathrm{R}} = U \sum_{\mathrm{cyc}} \left[ T(r_{ij}) \right]^2 \left[ T(r_{ik}) \right]^2, \tag{22}$$

with

$$x_{ij} = Y(r_{ij}) \sigma_i \cdot \sigma_j + T(r_{ij}) S_{\mathrm{T}ij}. \tag{23}$$

Correspondingly, the nuclear Hamiltonian caused by the UIX potential is separated into two parts:

$$H_3^{2\pi} = \sum_{i<j<k} V_{ijk}^{2\pi} . \tag{24}$$

$$H_3^{\mathrm{R}} = \sum_{i<j<k} V_{ijk}^{\mathrm{R}} . \tag{25}$$

In this paper, we evaluate the expectation values of $H_3^{2\pi}$ and $H_3^{\mathrm{R}}$ with the wave function for the Fermi gas at zero temperature, $\Phi_{\mathrm{F}}$. Explicit expressions of these expectation values per nucleon, $\langle H_3^{2\pi}\rangle_{\mathrm{F}}/N$ and $\langle H_3^{\mathrm{R}}\rangle_{\mathrm{F}}/N$ are

$$\frac{\langle H_3^{2\pi}\rangle_{\mathrm{F}}}{N} = -\frac{9A}{2\pi^2\rho}\int_0^\infty [G_{\mathrm{T2}}(k)]^2 G_{\mathrm{qL}}(k)k^2dk - \frac{9A}{4\pi^2\rho}\int_0^\infty [G_{\mathrm{Y}}(k)]^2 G_{\mathrm{qL}}(k)k^2dk$$
$$+\frac{9A}{8\pi^2\rho}\int_0^\infty [G_{\mathrm{TL2}}(k)]^2 G_{\mathrm{L}}(k)k^2dk + \frac{9A}{4\pi^2\rho}\int_0^\infty [G_{\mathrm{YL}}(k)]^2 G_{\mathrm{L}}(k)k^2dk , \tag{26}$$

$$\frac{\langle H_3^{\mathrm{R}}\rangle_{\mathrm{F}}}{N} = 8\pi^2 U\rho^2\left\{\int_0^\infty [T(r)]^2 r^2dr\right\}^2 - 4\pi^2 U\rho^2\int_0^\infty [T(r)]^2 r^2dr\int_0^\infty [T(r)]^2[l(k_{\mathrm{F}}r)]^2 r^2dr$$
$$-\frac{U}{16\pi^2\rho}\int_0^\infty [G_{\mathrm{qT}}(k)]^2 G_{\mathrm{qL}}(k)k^2dk + \frac{U}{32\pi^2\rho}\int_0^\infty [G_{\mathrm{qTL}}(k)]^2 G_{\mathrm{L}}(k)k^2dk, \tag{27}$$

with

$$G_{\mathrm{T2}}(k) = 4\pi\rho\int_0^\infty T(r)j_2(kr)r^2dr, \tag{28}$$

$$G_{\mathrm{qL}}(k) = 4\pi\rho\int_0^\infty [l(k_{\mathrm{F}}r)]^2 j_0(kr)r^2dr = \begin{cases} 4\left[1-\frac{3}{4}\frac{k}{k_{\mathrm{F}}}+\frac{1}{16}\left(\frac{k}{k_{\mathrm{F}}}\right)^3\right] & \text{for } k \le 2k_{\mathrm{F}} , \\ 0 & \text{for } k \ge 2k_{\mathrm{F}} , \end{cases} \tag{29}$$

$$G_{\mathrm{Y}}(k) = 4\pi\rho\int_0^\infty Y(r)j_0(kr)r^2dr, \tag{30}$$

$$G_{\mathrm{TL2}}(k) = 4\pi\rho\int_0^\infty T(r)l(k_{\mathrm{F}}r)j_2(kr)r^2dr, \tag{31}$$

$$G_{\rm L}(k)=4\pi\rho\int_0^\infty l(k_{\rm F}r)j_0(kr)r^2dr=\begin{cases}4 \ \text{ for } \ k<k_{\rm F}\ ,\\ 2 \ \text{ for } \ k=k_{\rm F}\ ,\\ 0 \ \text{ for } \ k>k_{\rm F}\ ,\end{cases} \tag{32}$$

$$G_{\rm YL}(k)=4\pi\rho\int_0^\infty Y(r)l(k_{\rm F}r)j_0(kr)r^2dr, \tag{33}$$

$$G_{\rm qT}(k)=4\pi\rho\int_0^\infty [T(r)]^2 j_0(kr)r^2dr, \tag{34}$$

$$G_{\rm qTL}(k)=4\pi\rho\int_0^\infty [T(r)]^2 l(k_{\rm F}r)j_0(kr)r^2dr, \tag{35}$$

where $l(z) = 3j_1(z)/z$. We note that, in the numerical calculations for Eqs. (26) and (27), we assume $G_{\rm L}(k = k_{\rm F}) = 4$, which does not affect the numerical results.

When we take into account the correlation between nucleons, the expectation values of $H_3^{2\pi}$ and $H_3^{\rm R}$ are likely to be different from $\langle H_3^{2\pi}\rangle_{\rm F}/N$ and $\langle H_3^{\rm R}\rangle_{\rm F}/N$, respectively. In addition, the possible relativistic boost correction will modify these nonrelativistic expectation values of TNI; in the FHNC calculation by APR, the expectation value of $H_3^{\rm R}$ was considerably reduced by this correction. Furthermore, in the case of symmetric nuclear matter, an additional term $\gamma\rho^2\exp[-\delta\rho]$ with $\gamma$ and $\delta$ being adjustable parameters was introduced so as to reproduce the empirical saturation point. In the present study, we take into account these corrections by introducing adjustable parameters $\alpha$, $\beta$, $\gamma$ and $\delta$, and express the correction energy $E_3/N$ as

$$\frac{E_3}{N}=\alpha\frac{\langle H_3^{\rm R}\rangle_{\rm F}}{N}+\beta\frac{\langle H_3^{2\pi}\rangle_{\rm F}}{N}+\gamma\rho^2e^{-\delta\rho}, \tag{36}$$

Then, the total energy per nucleon $E/N$ is

$$E/N = E_2/N + E_3/N. \tag{37}$$

In Eq. (36), the parameters $\alpha$, $\beta$, $\gamma$ and $\delta$ are assumed to be independent of $\rho$. Also, $\alpha$ and $\beta$ are common to symmetric nuclear matter and neutron matter. Here it is noted that $\langle H_3^{2\pi}\rangle_{\rm F}/N$ and $\langle H_3^{\rm R}\rangle_{\rm F}/N$ for symmetric nuclear matter are different from those for neutron matter because the corresponding Fermi-gas wave functions $\Phi_F$ are different from each other. The last term on the right-hand side of Eq. (36) is introduced only for symmetric nuclear matter, i.e., $\gamma = 0$ for neutron matter, as in the case of the FHNC calculations by APR. Then, we determine the values of the parameters so that the obtained $E/N$ reproduces the empirical saturation data, i.e., the saturation density $\rho_0$, saturation energy $E_0/N$, incompressibility $K$ and symmetry energy $E_s/N$. Here, we define the symmetry energy as the difference between $E/N$ for neutron matter and for symmetric nuclear matter at $\rho = \rho_0$, i.e., $E_s/N = E(\rho_0, 0)/N - E(\rho_0, 1/2)/N$, where $E(\rho, x)/N$ is the energy per nucleon for asymmetric nuclear matter at the density $\rho$ and the proton fraction $x = \rho_p/\rho$ ($\rho_p$ is the proton number density). The values of the parameters determined in this way are $\alpha$ = 0.41, $\beta = -0.22$, $\gamma = -1604$ MeV fm$^6$ and $\delta = 13.93$ fm$^3$, providing $\rho_0 = 0.16$ fm$^{-3}$, $E_0/N = -15.8$ MeV, $K = 250$ MeV, and $E_s/N = 30$ MeV.

The obtained $E/N$ for symmetric nuclear matter and neutron matter are compared with those by APR in Fig. 2. It is seen that, in this figure, $E/N$ for symmetric nuclear matter and neutron matter are in fair agreement with those by APR. We note that, in the APR calculation, the energy lowering occurs at $\rho \gtrsim 0.32$ fm$^{-3}$ (0.2 fm$^{-3}$) for symmetric nuclear matter (neutron matter) due to the $\pi^0$ condensation, while, in the present calculation, the $\pi^0$ condensation is not taken into account.

It is interesting to apply the obtained $E/N$ to neutron stars. Assuming $E(\rho, x)/N$ for asymmetric nuclear matter as

$$\frac{E}{N}(\rho,x) = \frac{E}{N}\left(\rho,\frac{1}{2}\right) + \left[\frac{E}{N}(\rho,0) - \frac{E}{N}\left(\rho,\frac{1}{2}\right)\right](1-2x)^2, \tag{38}$$

we evaluate the energy per nucleon $E/N$ for neutron star matter as the charge-neutral beta-stable mixture of nucleons, electrons and, if any, muons with the leptons being treated as the relativistic Fermi gases. Then, we solve the TOV equation to obtain the neutron star structure. In Fig. 3, the obtained masses of the neutron stars are plotted as a function of the central matter density $\rho_{m0}$. The maximum mass is 2.22 $M_\odot$. Here, we should note that, at high densities, the causality violation occurs, i.e., the sound velocity $v_s$ exceeds the speed of light $c$. In order to see this, we plot $v_s/c$ as a function of the matter density $\rho_m$ in Fig. 4. This figure shows that the causality is violated at the matter density higher than the critical density $\rho_{mc}$ ~ $1.91\times10^{15}$ g/cm$^3$. Therefore, in Fig. 3, only the solutions in the region $\rho_{m0} < \rho_{mc}$ are valid, and the maximum mass of the neutron star is 2.16 $M_\odot$ in this region.

3. Nuclear matter at finite temperatures

In this section, we calculate the free energy per nucleon $F/N$ for nuclear matter at finite temperatures using the prescription proposed by Schmidt and Pandharipande (SP). We start from the following expression,

$$\frac{F}{N} = \frac{E_{0T}}{N} - T\frac{S_0}{N}, \tag{39}$$

where $E_{0T}/N$ is the approximate internal energy per nucleon, $S_0/N$ is the approximate entropy per nucleon and $T$ is the temperature. The approximate internal energy $E_{0T}/N$ is expressed as,

$$\frac{E_{0T}}{N} = \frac{E_{2T}}{N} + \frac{E_{3T}}{N}. \tag{40}$$

Here, $E_{2T}$ is the expectation value of $H_2$ at finite temperature evaluated as follows. First, we introduce the Jastrow-type wave function at finite temperature

$$\psi(T) = \mathrm{Sym}\left[\prod_{i<j} f_{ij}\right]\Phi_{F}(T), \tag{41}$$

where $f_{ij}$ is the two-body correlation function defined as in Eq. (4), and $\Phi_F(T)$ is the Fermi-gas wave function at finite temperature specified by the average occupation probability $n(k)$ of the quasi-particle states:

$$n(k) = \left\{1 + \exp[\frac{\varepsilon(k) - \mu_0}{k_B T}]\right\}^{-1}. \tag{42}$$

In Eq. (42), $k_B$ is the Boltzmann constant and $\varepsilon(k)$ is the quasi-particle energy which is assumed as

$$\varepsilon(k) = \frac{\hbar^2 k^2}{2m^*}, \tag{43}$$

with $m^*$ being the effective mass of the quasi-particle. The approximate chemical potential $\mu_0$ in Eq. (42) is determined by the condition,

$$\frac{1}{\Omega}\sum_i n(k) = \frac{2}{\pi^2}\int_0^\infty n(k)k^2 dk = \rho. \tag{44}$$

Using the wave function at finite temperature Eq. (41), we evaluate $E_{2T}$ as the two-body cluster approximation of $\langle H_2 \rangle_T = \langle \Psi(T)|H_2|\Psi(T)\rangle / \langle \Psi(T)|\Psi(T)\rangle$. The explicit expression for $E_{2T}/N$ is the same as for $E_2/N$ at zero temperature given in Eq. (5) except that the one-body kinetic energy $3/5E_F$ is replaced by

$$\frac{E_{1T}}{N} = \frac{\hbar^2}{m}\frac{1}{\pi^2\rho}\int_0^\infty n(k)k^4 dk, \tag{45}$$

and that $F_{Fts}(r)$, $F_{qFts}(r)$ and $F_{bFts}(r)$ are replaced by $F_{Fts}(r;T)$, $F_{qFts}(r;T)$ and $F_{bFts}(r;T)$ defined as

$$F_{Fts}(r;T) = \frac{(2t+1)(2s+1)}{16}\left\{1-(-1)^{t+s}\left[\frac{2}{\pi^2\rho}\int_0^\infty n(k)j_0(kr)k^2 dk\right]^2\right\}, \tag{46}$$

$$F_{qFts}(r;T) = \frac{(2t+1)(2s+1)}{16}\left[\frac{2}{3\pi^2\rho}\int_0^\infty n(k)k^4 r^2 dk \right.$$
$$\left. +(-1)^{t+s}\left(\frac{2}{\pi^2\rho}\right)^2\int_0^\infty n(k)j_0(kr)k^2 dk\int_0^\infty n(k')j_1(k'r)rk'^3 dk'\right], \tag{47}$$

$$F_{bFts}(r;T) = \frac{(2t+1)(2s+1)}{16}\left\{\frac{2}{\pi^2\rho}\int_0^\infty n(k)\left(\frac{2}{3}k^4r^2 + \frac{1}{15}k^6r^4\right)dk\right.$$

$$+\left[\frac{2}{3\pi^2\rho}\int_0^\infty n(k)k^4r^2dk\right]^2-(-1)^{t+s}\left\{\left[\frac{2}{\pi^2\rho}\int_0^\infty n(k)j_1(kr)k^2dk\right]^2\right.$$

$$\left.-\frac{2}{\pi^2\rho}\int_0^\infty n(k)j_0(kr)k^2dk\frac{1}{\pi^2\rho}\int_0^\infty n(k')\left[j_0(k'r)r^2k'^4-j_1(k'r)rk'^3\right]dk'\right\}. \quad (48)$$

Following the procedure of SP and FP, we simply assume that the correlation functions $f_{Cts}(r)$, $f_{Tt}(r)$ and $f_{SOt}(r)$ in the expression for $E_{2T}/N$ are the same as those at zero temperature determined by the variational calculation explained in the previous section. Here, we note that the bare nucleon mass is used in the kinetic-energy operator for $E_{2T}/N$; the effective mass $m^*$ appears only in $n(k)$.

The last term on the right-hand side of Eq. (40), $E_{3T}/N$, represents the correction to the approximated internal energy at finite temperature caused by TNI. In this study, we simply approximate that $E_{3T}/N$ is the same as $E_3/N$ at zero temperature shown in Eq. (36), i.e., we neglect the thermal effect on it. The validity of this approximation is discussed later.

Finally, the approximate entropy per nucleon $S_0/N$ in Eq. (39) is expressed in terms of $n(k)$ as,

$$\frac{S_0}{N}=-\frac{k_B}{N}\sum_i\{[1-n(k)]\ln[1-n(k)]+n(k)\ln n(k)\}$$

$$=-\frac{2k_B}{\pi^2\rho}\int_0^\infty\{[1-n(k)]\ln[1-n(k)]+n(k)\ln n(k)\}k^2dk. \quad (49)$$

As a result, $F/N$ is a functional of $n(k)$, or, more specifically, a function of the effective mass $m^*$. Then, $F/N$ is minimized with respect to $m^*$ under the normalization condition Eq. (44). We note that Eqs. (44)~(49) are for symmetric nuclear matter; corresponding expressions for neutron matter are derived similarly.

The obtained $F/N$ for symmetric nuclear matter and neutron matter are plotted in Fig. 5 compared with those by FP. It is seen that, at low densities, $F/N$ is close to that of FP-EOS for both symmetric nuclear matter and neutron matter. On the other hand, in the high-density region, $F/N$ for symmetric nuclear matter and neutron matter are higher than those by FP which means that the EOS constructed in this paper is stiffer than the FP-EOS. In fact, the maximum mass of the cold neutron star using the FP-EOS is just below 2 $M_\odot$, which is smaller than 2.22 $M_\odot$ in the present study.

In the low-density region, the pressure $P$ derived from $F/N$ for symmetric nuclear matter is shown in Fig. 6 together with that by FP. The pressure in the present study is very close to that by FP at zero temperature, but it is somewhat lower than that by FP at $T \gtrsim 10$ MeV. The critical temperature $T_c$ is estimated to be about 18 MeV, which is reasonable. At higher densities, the pressure in the present study is considerably higher than that by FP.

The internal energy per nucleon $E_T/N$ and the entropy per nucleon $S/N$ are shown in Figs. 7 and 8, respectively. Here, $E_T/N$ and $S/N$ are obtained with thermodynamic relations as follows:

$$\frac{E_T}{N} = \frac{F}{N} + T\frac{S}{N}, \tag{50}$$

$$\frac{S}{N} = -\frac{\partial}{\partial T}\frac{F}{N}\bigg|_\rho . \tag{51}$$

In Fig. 7, the approximate internal energy $E_{0T}/N$ is also plotted. It is noted that, as seen in Fig. 7, $E_T/N$ is in good agreement with $E_{0T}/N$. Correspondingly, $S/N$ is also in good agreement with the approximate entropy $S_0/N$. This self-consistency implies the validity of the present prescription for the nuclear EOS at finite temperatures. In Fig. 8, $S/N$ is compared with that by FP. The entropy in the present study is close to that by FP; in the high-density region, the present $S/N$ is slightly higher.

The effective mass $m^*$, which is the variational parameter, is given in Fig. 9. According to SP, $m^*$ is regarded as the effective mass of the quasi-particle at $k = k_F$. In fact, the value of $m^*$ for symmetric nuclear matter at the normal density is ~ 0.7 $m$, which is fairly reasonable. Compared with the results by FP, the density dependence of $m^*$ in the present calculations is rather moderate: At higher densities, $m^*$ is larger than that by FP.

We discuss here the thermal energy $\Delta E_T/N$ defined as

$$\frac{\Delta E_{\mathrm{T}}}{N} = \frac{E_{0\mathrm{T}}}{N} - \frac{E}{N}. \tag{52}$$

Strictly speaking, $E_{\mathrm{T}}/N$ should be used in Eq. (52) instead of $E_{0\mathrm{T}}/N$, but, as pointed out above, $E_{0\mathrm{T}}/N$ is in good agreement with $E_{\mathrm{T}}/N$. In Fig. 10, $\Delta E_{\mathrm{T}}/N$ for symmetric nuclear matter and neutron matter are shown and compared with those by FP. It is seen that $\Delta E_{\mathrm{T}}/N$ is slightly higher than those of FP-EOS in the high-density region. In order to study the thermal energy more in detail, we decompose $\Delta E_{\mathrm{T}}/N$ as

$$\frac{\Delta E_{\mathrm{T}}}{N} = \frac{\Delta E_{1}}{N} + \frac{\Delta E_{2\mathrm{qL}}}{N} + \frac{\Delta E_{2\mathrm{r}}}{N}. \tag{53}$$

Here, $\Delta E_1/N = E_{1\mathrm{T}}/N - 3/5\, E_{\mathrm{F}}$ is the thermal energy caused by the one-body kinetic energy. The second term $\Delta E_{2\mathrm{qL}}/N$ represents the thermal energy caused by the part of the two-body potential that is proportional to the quadratic $\boldsymbol{L}_{ij}$, i.e., the expectation values of the part of the potential energy including $V_{\mathrm{qL}ts}(r)$ and $V_{\mathrm{qSO}t}(r)$ minus the corresponding quantities at zero temperature. The last term $\Delta E_{2\mathrm{r}}/N$ is the remaining part of the thermal energy composed of the central, tensor and spin-orbit potential energy and the kinetic energy caused by the correlation.

In Fig. 11, $\Delta E_1/N$, $\Delta E_{2\mathrm{qL}}/N$ and $\Delta E_{2\mathrm{r}}/N$ are plotted in the case of symmetric nuclear matter at $T$ = 30 MeV. This figure shows that $\Delta E_1/N$ is dominant in the regions of relatively low density. At higher densities, $\Delta E_{2\mathrm{qL}}/N$ becomes comparable to $\Delta E_1/N$, but the contribution of $\Delta E_{2\mathrm{r}}/N$ is still small. This implies that the expectation value of the operator including quadratic $\boldsymbol{L}$ ($\Delta E_{2\mathrm{qL}}/N$) or quadratic momentum (the one-body kinetic energy term $\Delta E_1/N$) is sensitive to the temperature, while the major two-body term $\Delta E_{2\mathrm{r}}/N$ is not. This property is deeply related to the normalization condition Eq. (44). In fact, as shown in Eq. (46), the direct term of $F_{\mathrm{F}ts}(r;\, T)$ is independent of $n(k)$; the temperature dependence appears only in the exchange term of $F_{\mathrm{F}ts}(r;\, T)$.

Similar considerations will hold for the correction energy $E_{3\mathrm{T}}/N$ caused mainly by TNI. In this paper, we assume that $E_{3\mathrm{T}}/N$ is the same as $E_3/N$ at zero temperature, which means that the thermal energy in $E_{3\mathrm{T}}/N$ is assumed to be small. From the considerations as above, it is expected that the assumption is valid because the UIX potential does not include quadratic-***L*** terms.

## 4. Concluding remarks

In this paper, we constructed an EOS for uniform nuclear matter by the variational method. The obtained EOS is suitable for constructing a new nuclear EOS for supernova simulations for the following reasons: (i) Since the present calculation is based on the realistic nuclear Hamiltonian, the EOS is directly connected with the bare nuclear forces, as desired in the studies of supernovae. (ii) The nuclear EOS at zero temperature and at finite temperatures are constructed in a consistent way because the relatively simple two-body cluster approximation is employed. The direct calculation of the energy for asymmetric nuclear matter instead of the interpolation Eq. (38), which is the next task to construct an EOS for supernovae, is also possible within this approximation. (iii) At zero temperature, the EOS in the present study is in good agreement with that by APR, owing to the healing-distance condition. (iv) The EOS reproduces the empirical saturation conditions at zero temperature. This reproduction is essential for the next step of our research. At densities lower than the normal density, nucleon clusters are created, and then nuclear matter becomes nonuniform. In order to describe the nucleon clusters in the Thomas-Fermi approximation with use of the EOS for uniform nuclear matter as obtained in this paper, it is crucial whether the empirical saturation conditions are reproduced or not.

Finally, we comment on the possible phase transitions in uniform nuclear matter at high densities. Various transitions to exotic phases such as the pion condensations, hyperon mixing and hadron-quark phase transition, have been proposed, and it is interesting to study their influences on supernovae. In fact, the FHNC calculation by APR includes the pion condensation automatically as explained above. The nuclear EOS with the hyperon mixing is studied based on the many-body theories, e.g., in [14]. Further studies to include the exotic phases in the current approach are important, and remain to be our future problem.

## Acknowledgements

We would like to express special thanks to Profs. S. Yamada, M. Yamada and K. Iida for valuable discussions. We also appreciate the cooperation of Messrs. K. Tanaka and K. Murai received in the early stage of this study. This study is supported by a Grant-in-Aid for the 21st century COE program “Holistic Research and Education Center for Physics of Self-organizing Systems” at Waseda University, and Grants-in-Aid from the Scientific Research Fund of the JSPS (Nos. 15540243, 18540291 and 18540295). Some of the numerical calculations were performed with a SR8000/MPP at the Information Technology Center of the University of Tokyo.

Figure captions

Fig. 1 Energies per nucleon without TNI, $E_2/N$, for symmetric nuclear matter and neutron matter as functions of the density. Corresponding $\langle H_2 \rangle/N$ by APR are also shown.

Fig. 2 Energies per nucleon including TNI, $E/N$, for symmetric nuclear matter and neutron matter as functions of the density. The energies obtained by APR are also shown.

Fig. 3 Neutron star mass as a function of the central matter density. The vertical line shows the critical density above which the causality is violated.

Fig. 4 Sound velocity of the neutron star matter as a function of the matter density.

Fig. 5 Free energies per nucleon, $F/N$, at $T$ = 0, 5, 10, 15, 20, 25 and 30 MeV for symmetric nuclear matter (a) and neutron matter (b) as functions of the density. The free energies obtained by FP at $T$ = 0, 10 and 20 MeV for symmetric nuclear matter and neutron matter are also shown.

Fig. 6 Pressures for symmetric nuclear matter at $T$ = 0, 5, 10, 15, 20, 25 and 30 MeV as functions of the density. The pressures obtained by FP at $T$ = 0, 5, 10, 15 and 20 MeV are also shown.

Fig. 7 Internal energies per nucleon, $E_T/N$, derived from $F/N$ at $T$ = 0, 5, 10, 15, 20 and 25 MeV for symmetric nuclear matter (a) and neutron matter (b) as functions of the density. The approximate internal energies per nucleon $E_{0T}/N$ are also shown.

Fig. 8 Entropies per nucleon $S/N$ derived from $F/N$ at $T$ = 5, 10, 15, 20 and 25 MeV for symmetric nuclear matter (a) and neutron matter (b) as functions of the density. For comparison, the entropies per nucleon obtained by FP at $T$ = 10 and 20 MeV for symmetric nuclear matter and neutron matter are also plotted.

Fig. 9 Effective masses $m^*$ at $T$ = 5 and 20 MeV for symmetric nuclear matter and neutron matter as functions of the density. The effective masses obtained by FP at $T$ = 20 MeV for symmetric nuclear matter and neutron matter are also plotted.

Fig. 10 Thermal energies per nucleon $\Delta E_{\mathrm{T}}/N$ at $T$ = 5, 10, 15, 20, 25 and 30 MeV for symmetric nuclear matter (a) and neutron matter (b) as functions of the density. For comparison, the thermal energies per nucleon obtained by FP at $T$ = 10 and 20 MeV for symmetric nuclear matter and neutron matter are also plotted.

Fig. 11 Three components of the thermal energy, $\Delta E_1/N$, $\Delta E_{2\mathrm{qL}}/N$ and $\Delta E_{2\mathrm{r}}/N$ for symmetric nuclear matter at $T$ = 30 MeV as functions of the density.

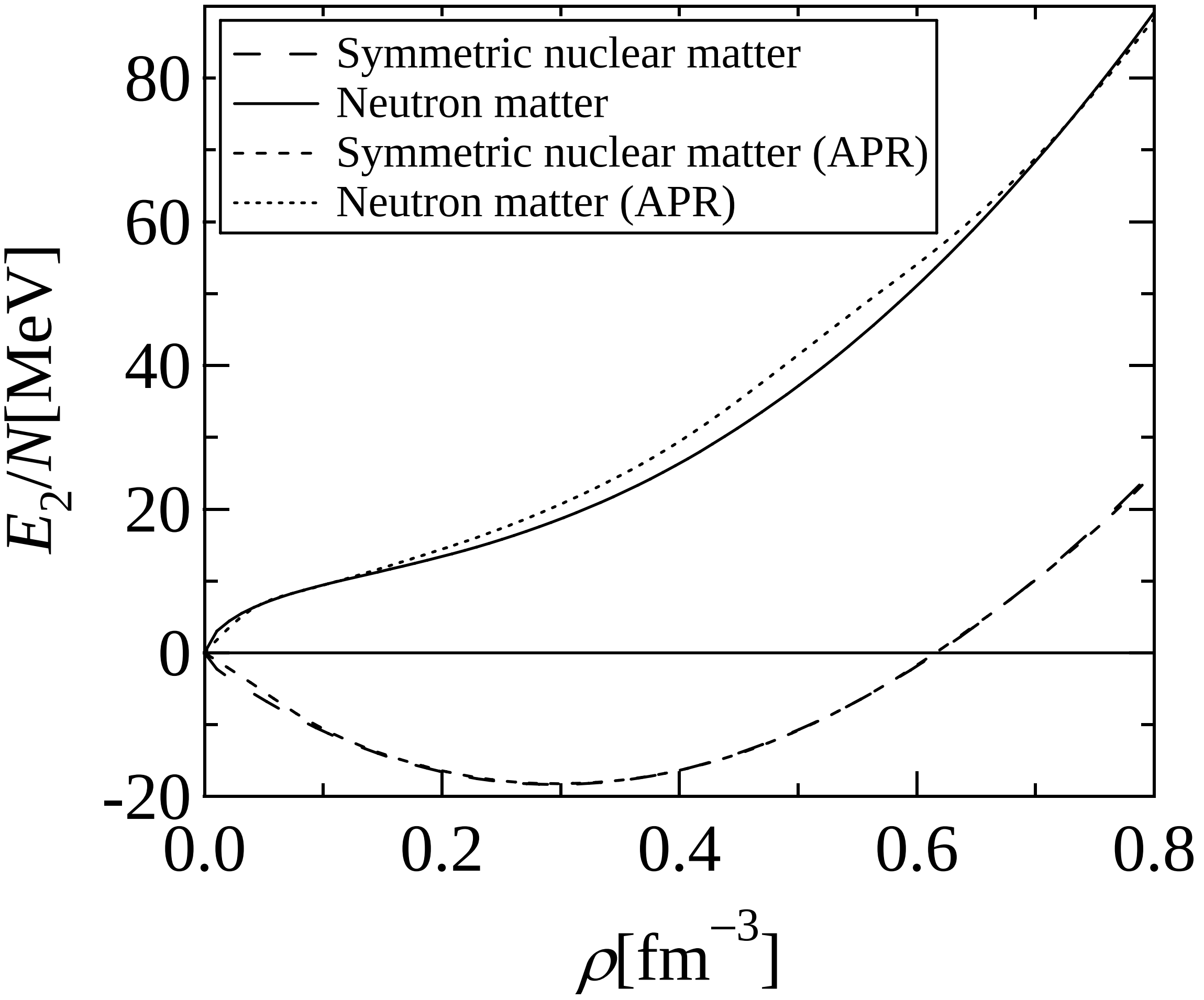

Fig.1

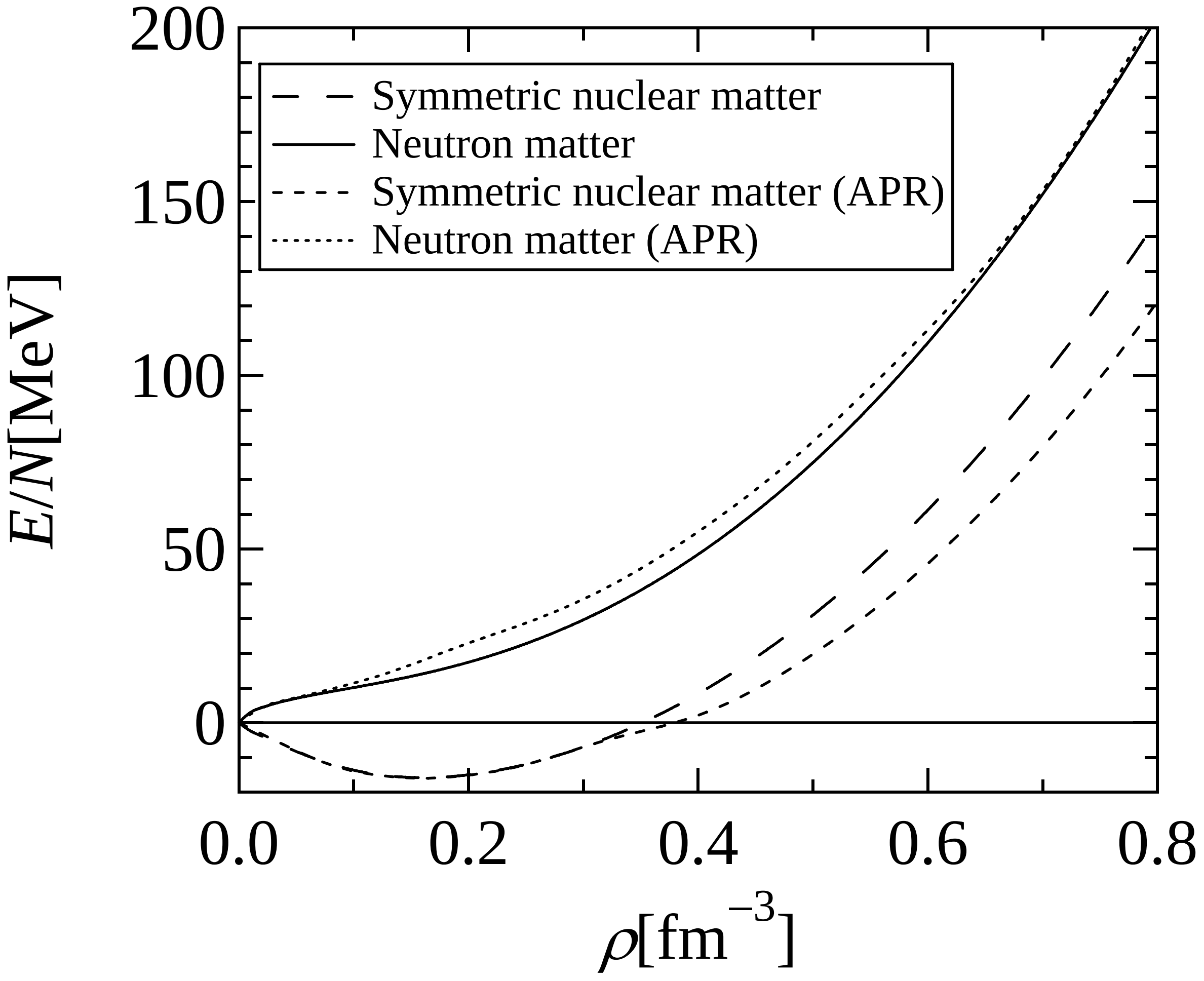

Fig.2

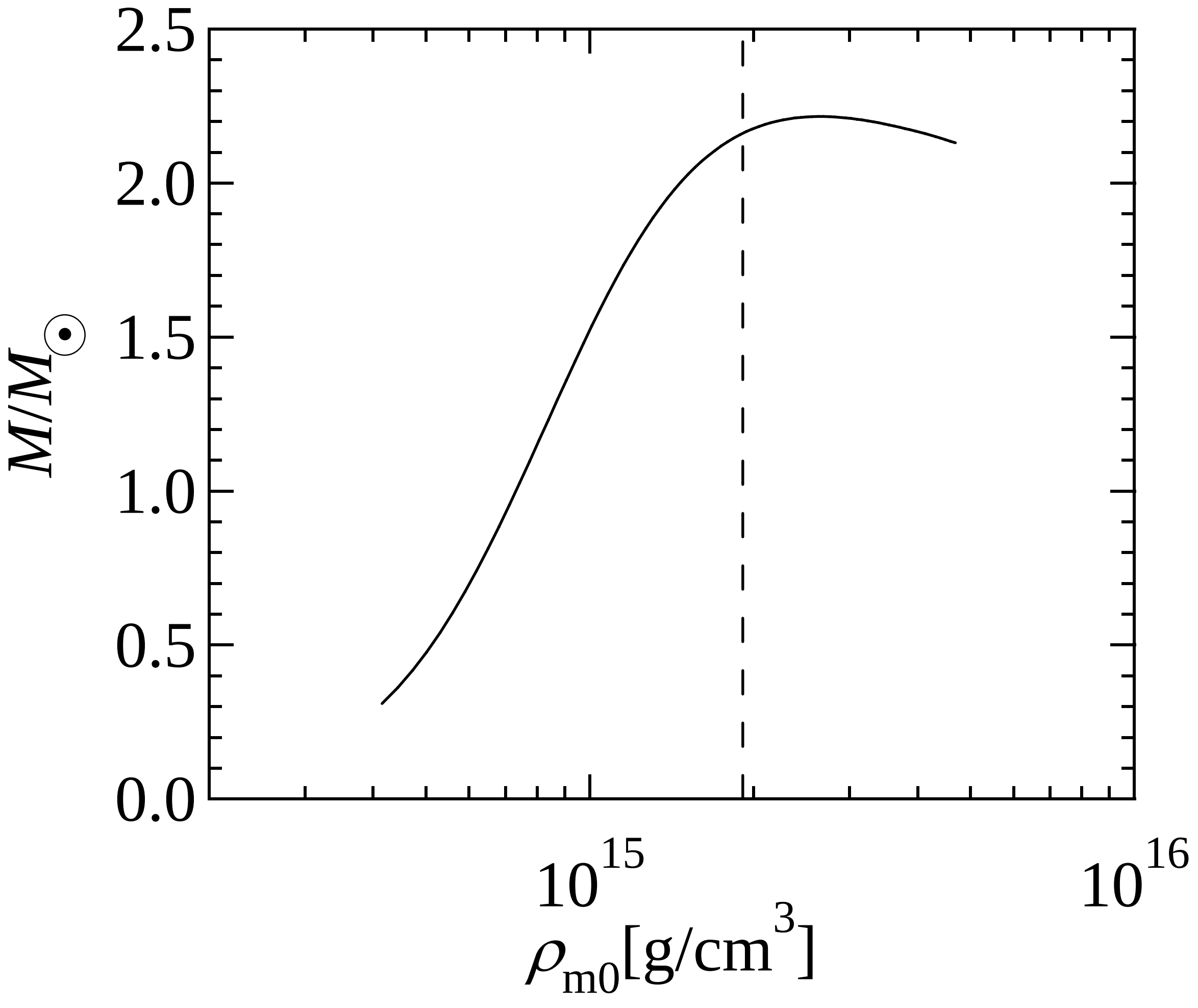

Fig.3

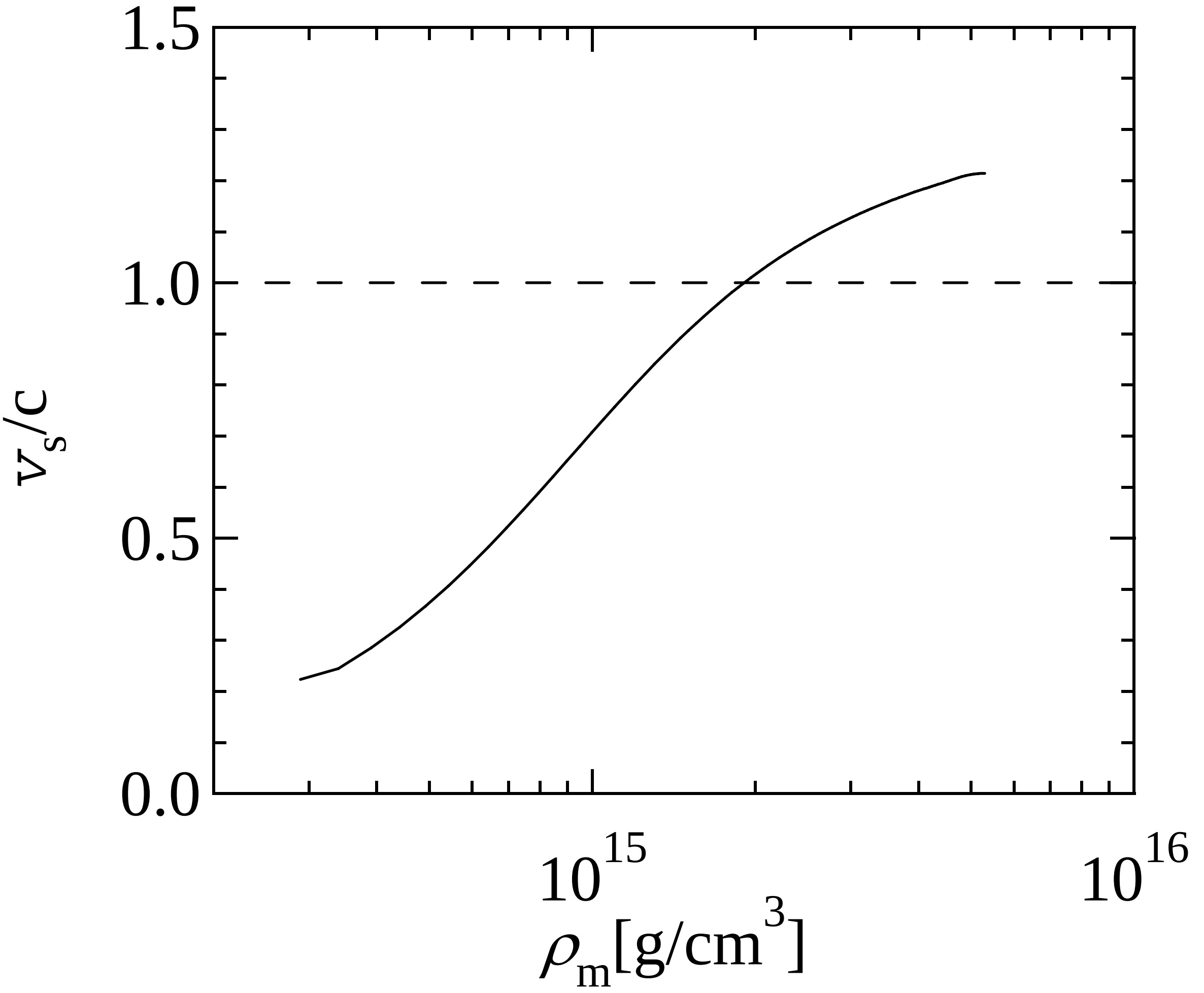

Fig.4

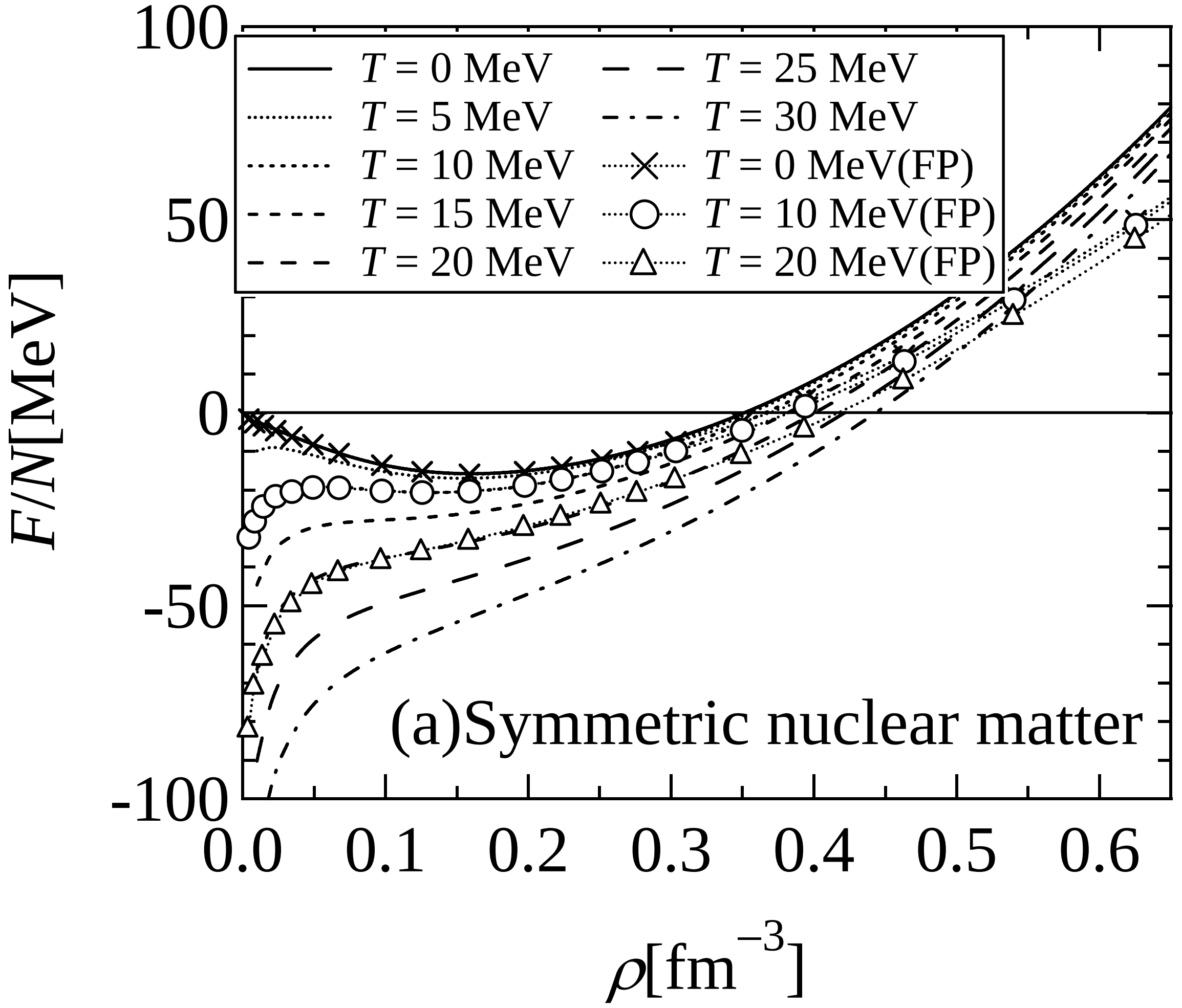
100
50
0
-50
-100
0.0
0.1
0.2
0.3
0.4
0.5
0.6
F/N[MeV]
ρ[fm^−3]
T = 0 MeV
T = 5 MeV
T = 10 MeV
T = 15 MeV
T = 20 MeV
T = 25 MeV
T = 30 MeV
T = 0 MeV(FP)
T = 10 MeV(FP)
T = 20 MeV(FP)
(a)Symmetric nuclear matter

Fig5(a)

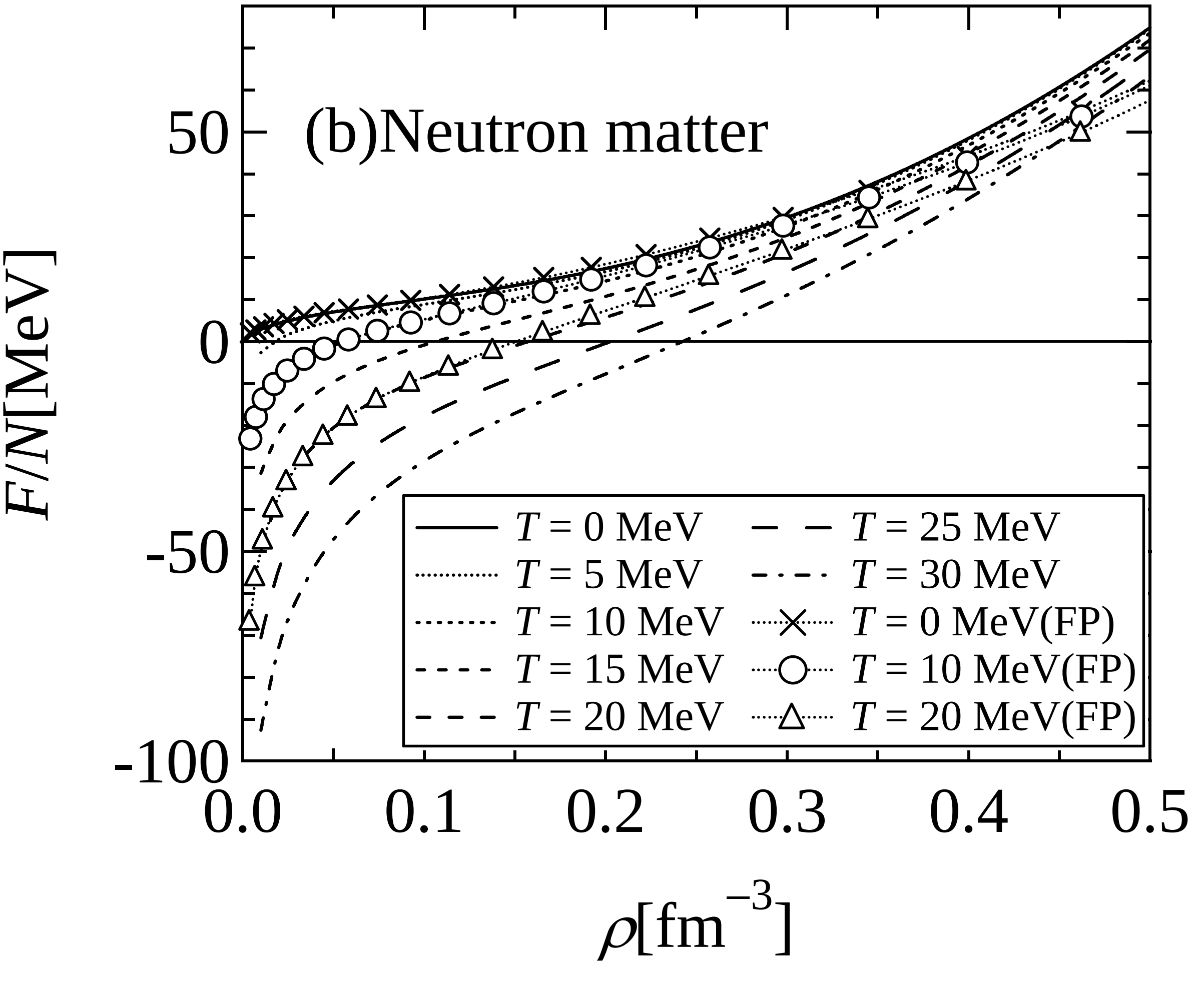
50
0
-50
-100
0.0
0.1
0.2
0.3
0.4
0.5
F/N[MeV]
ρ[fm^−3]
(b)Neutron matter
T = 0 MeV
T = 5 MeV
T = 10 MeV
T = 15 MeV
T = 20 MeV
T = 25 MeV
T = 30 MeV
T = 0 MeV(FP)
T = 10 MeV(FP)
T = 20 MeV(FP)

Fig5(b)

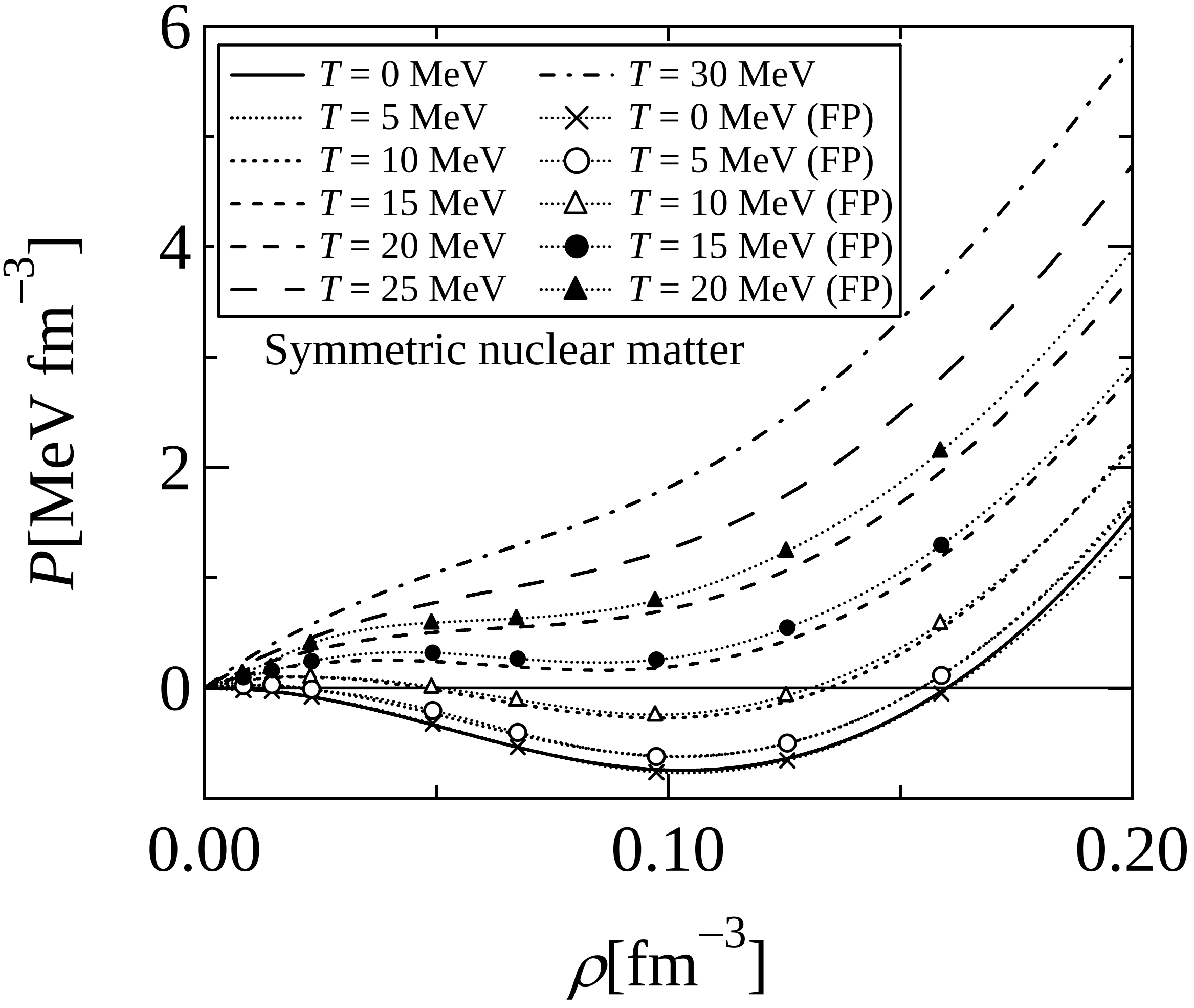

T = 0 MeV
T = 5 MeV
T = 10 MeV
T = 15 MeV
T = 20 MeV
T = 25 MeV
T = 30 MeV
T = 0 MeV (FP)
T = 5 MeV (FP)
T = 10 MeV (FP)
T = 15 MeV (FP)
T = 20 MeV (FP)
Symmetric nuclear matter
P[MeV fm⁻³]
ρ[fm⁻³]
0.00
0.10
0.20
0
2
4
6

Fig. 6

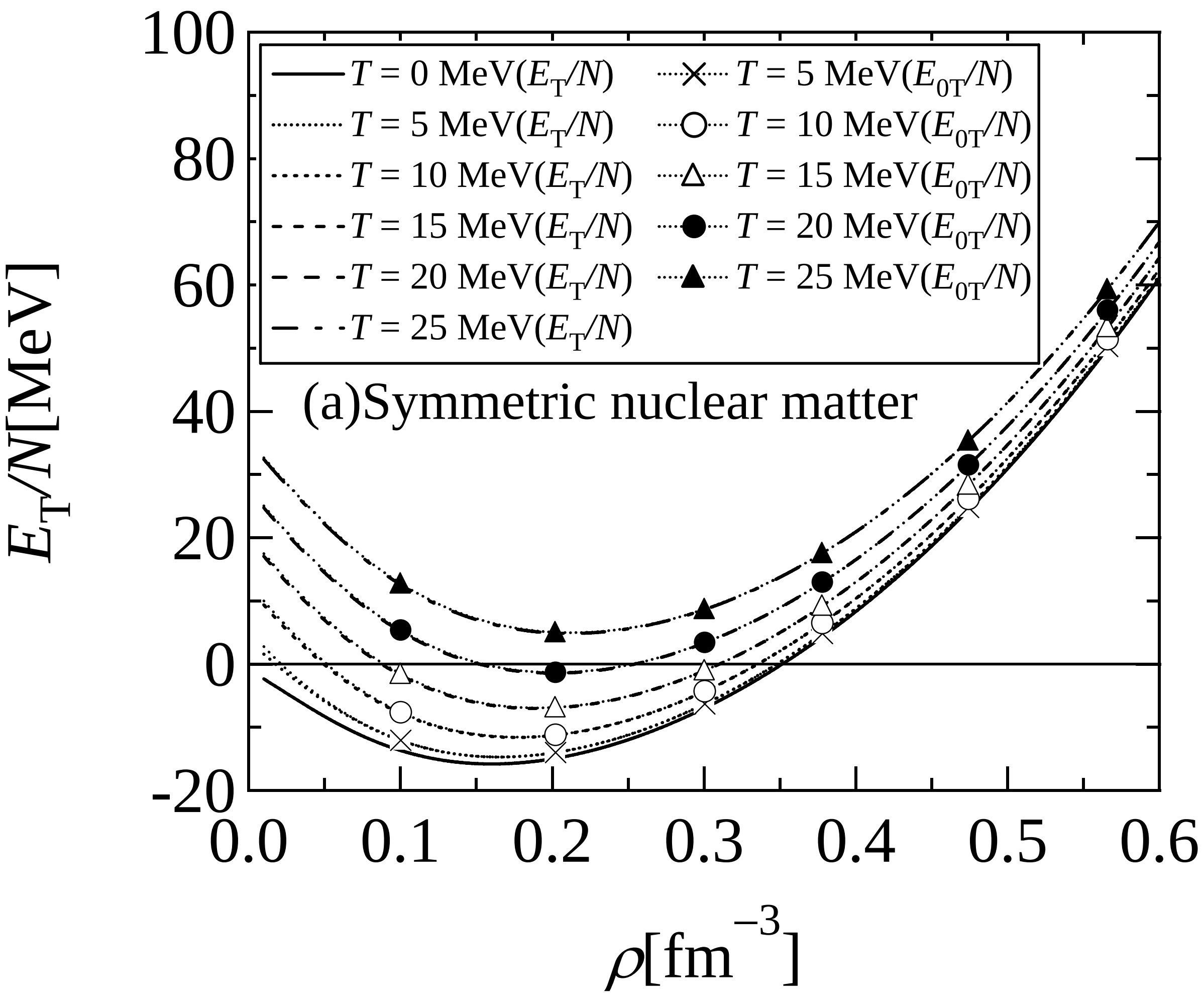

T = 0 MeV(E_T/N)
T = 5 MeV(E_T/N)
T = 10 MeV(E_T/N)
T = 15 MeV(E_T/N)
T = 20 MeV(E_T/N)
T = 25 MeV(E_T/N)
T = 5 MeV(E_0T/N)
T = 10 MeV(E_0T/N)
T = 15 MeV(E_0T/N)
T = 20 MeV(E_0T/N)
T = 25 MeV(E_0T/N)
(a)Symmetric nuclear matter
E_T/N[MeV]
ρ[fm^−3]
100
80
60
40
20
0
-20
0.0
0.1
0.2
0.3
0.4
0.5
0.6

Fig. 7(a)

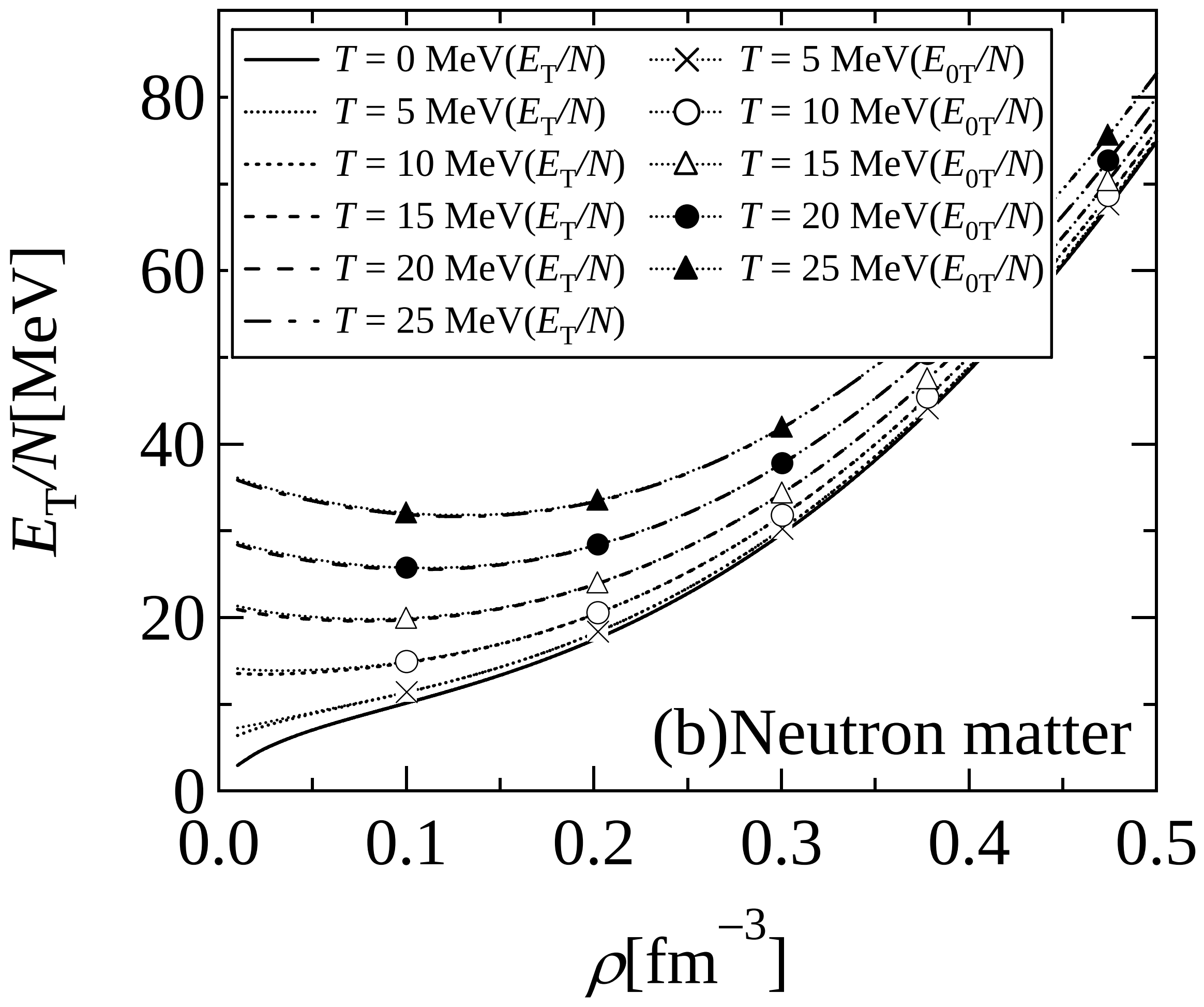

T = 0 MeV(E_T/N)
T = 5 MeV(E_T/N)
T = 10 MeV(E_T/N)
T = 15 MeV(E_T/N)
T = 20 MeV(E_T/N)
T = 25 MeV(E_T/N)
T = 5 MeV(E_0T/N)
T = 10 MeV(E_0T/N)
T = 15 MeV(E_0T/N)
T = 20 MeV(E_0T/N)
T = 25 MeV(E_0T/N)
(b)Neutron matter
E_T/N[MeV]
ρ[fm^−3]
80
60
40
20
0
0.0
0.1
0.2
0.3
0.4
0.5

Fig. 7(b)

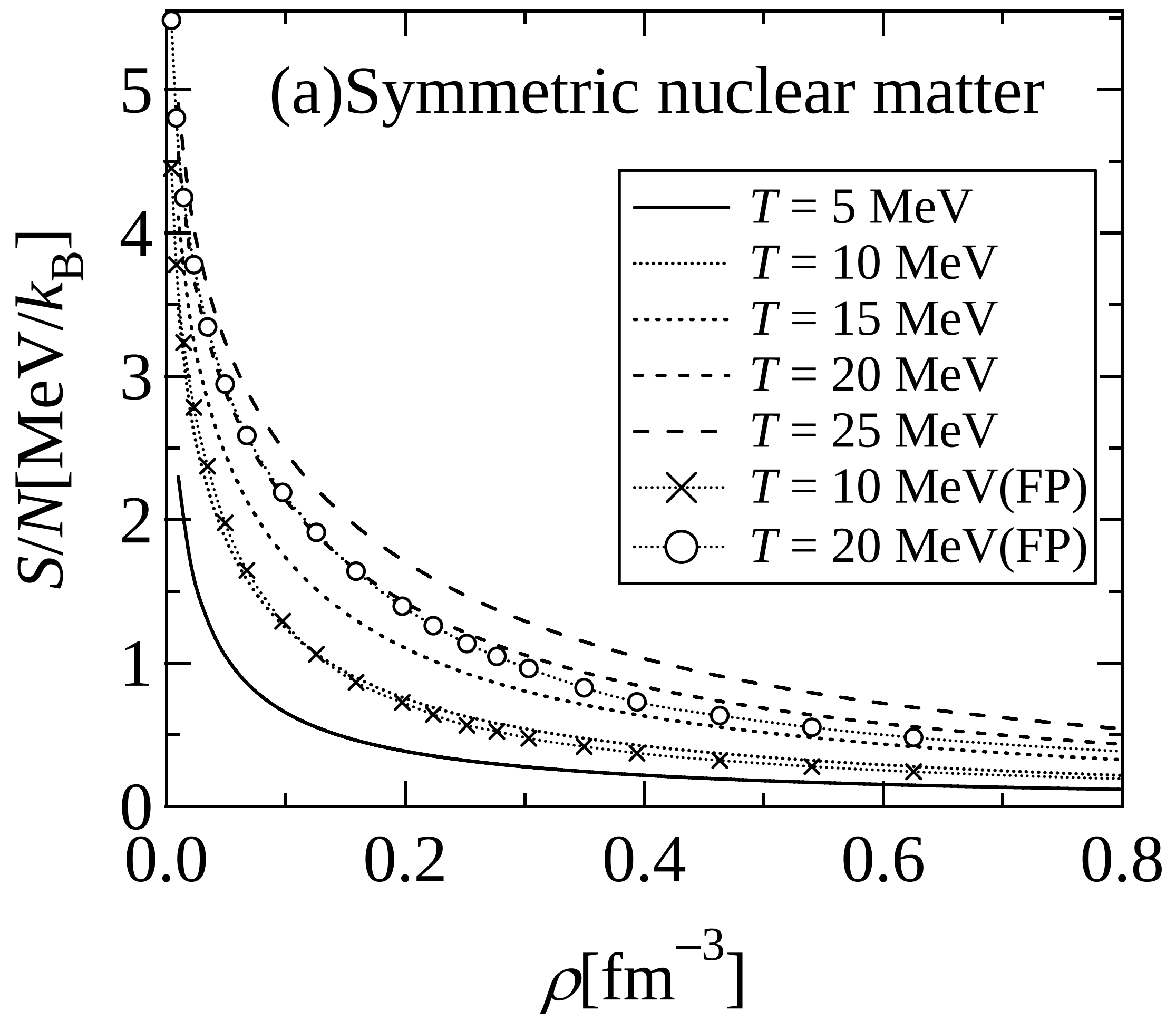
(a)Symmetric nuclear matter
T = 5 MeV
T = 10 MeV
T = 15 MeV
T = 20 MeV
T = 25 MeV
T = 10 MeV(FP)
T = 20 MeV(FP)
S/N[MeV/k_B]
ρ[fm^-3]
0
1
2
3
4
5
0.0
0.2
0.4
0.6
0.8

Fig.8(a)

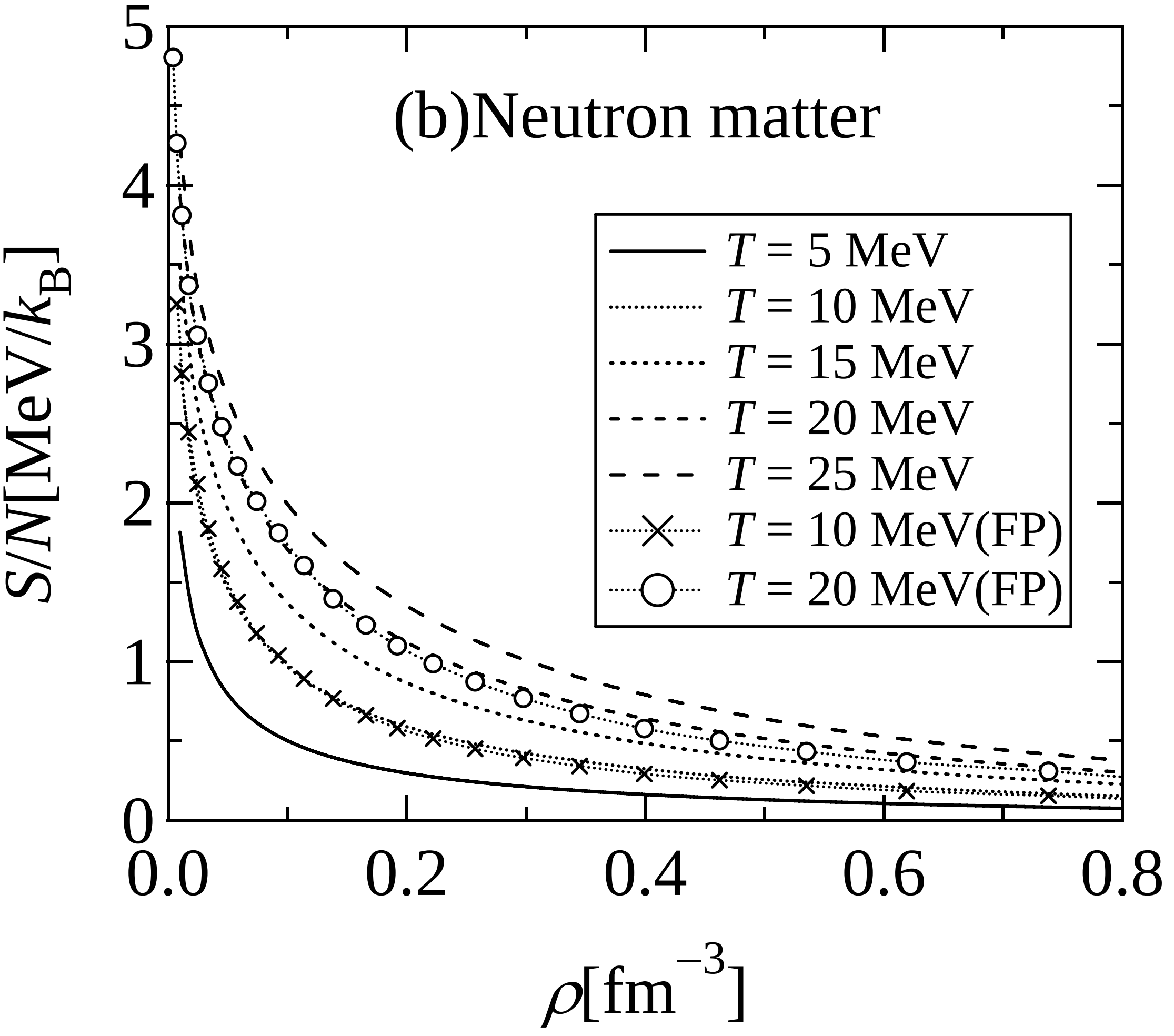
(b)Neutron matter
T = 5 MeV
T = 10 MeV
T = 15 MeV
T = 20 MeV
T = 25 MeV
T = 10 MeV(FP)
T = 20 MeV(FP)
S/N[MeV/k_B]
ρ[fm^-3]
0
1
2
3
4
5
0.0
0.2
0.4
0.6
0.8

Fig.8(b)

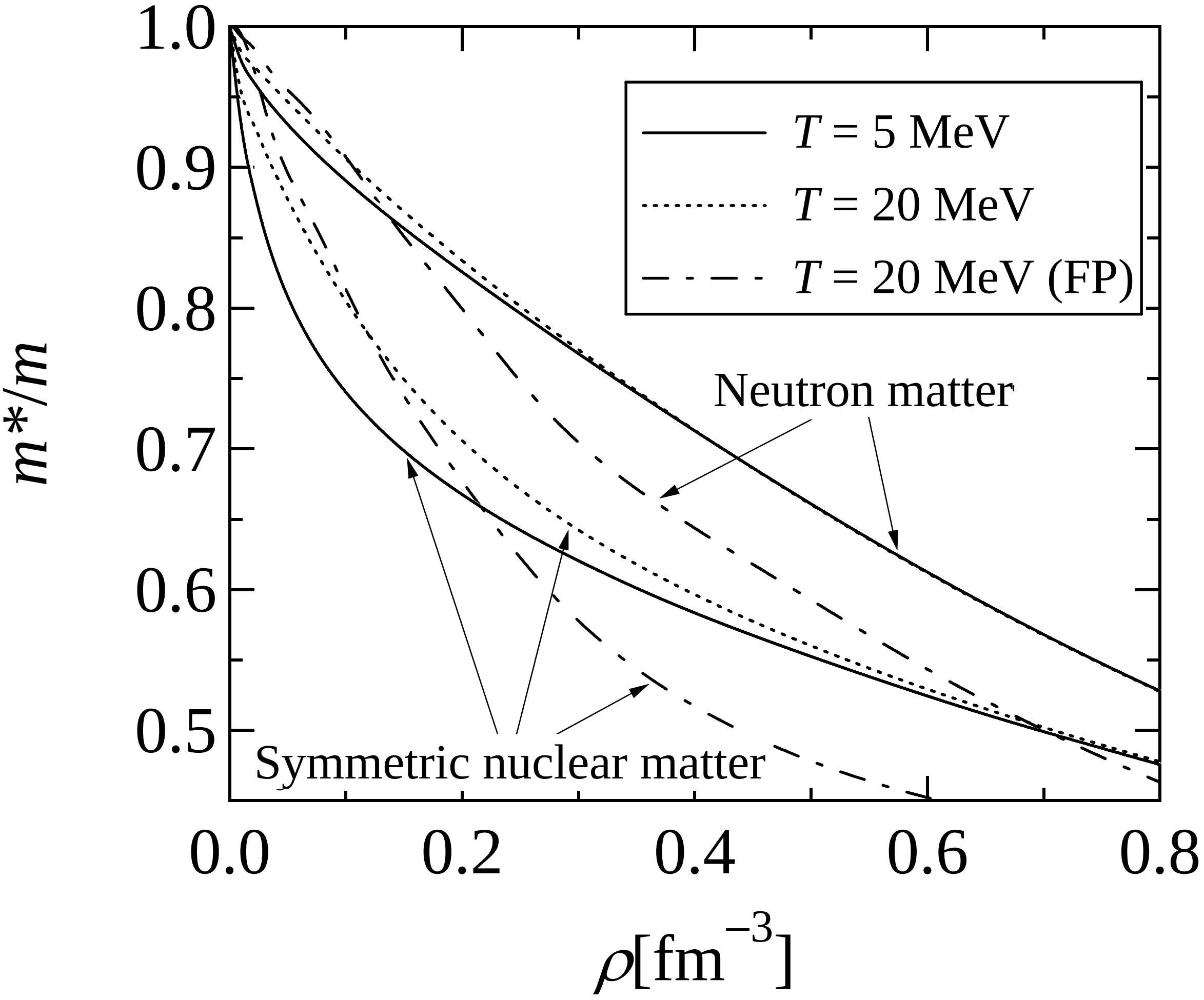

Fig. 9

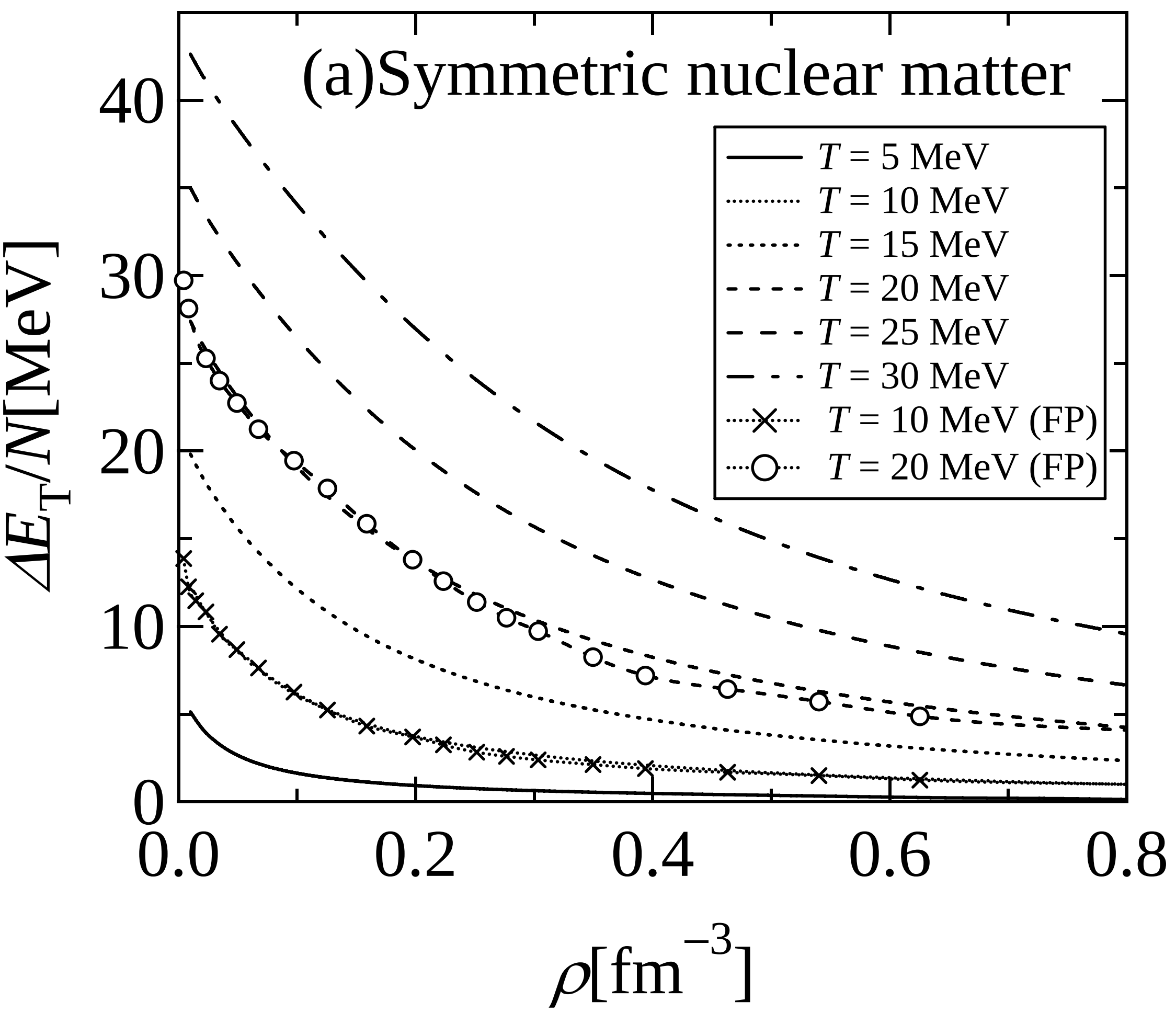

Fig.10(a)

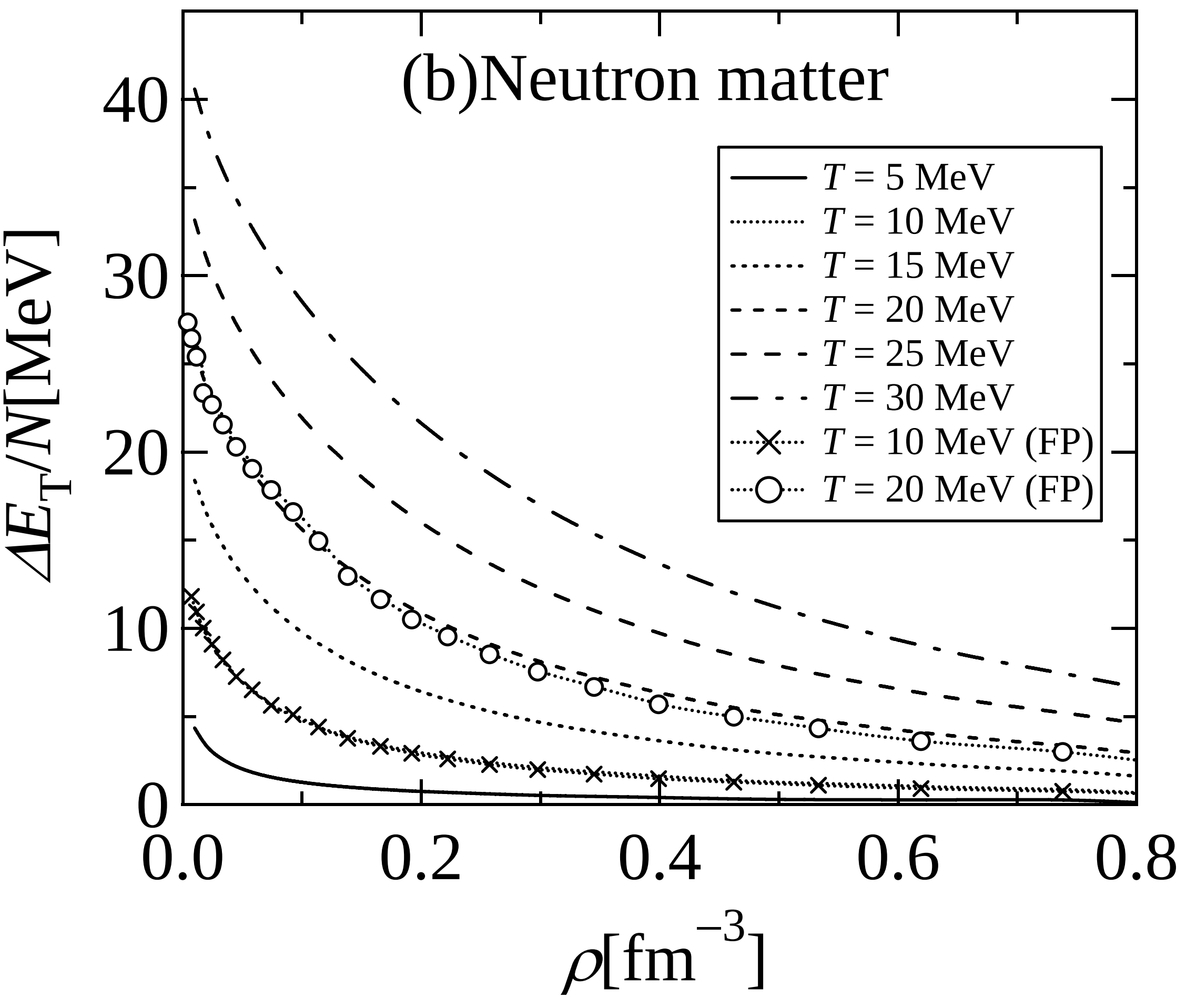

Fig.10(b)

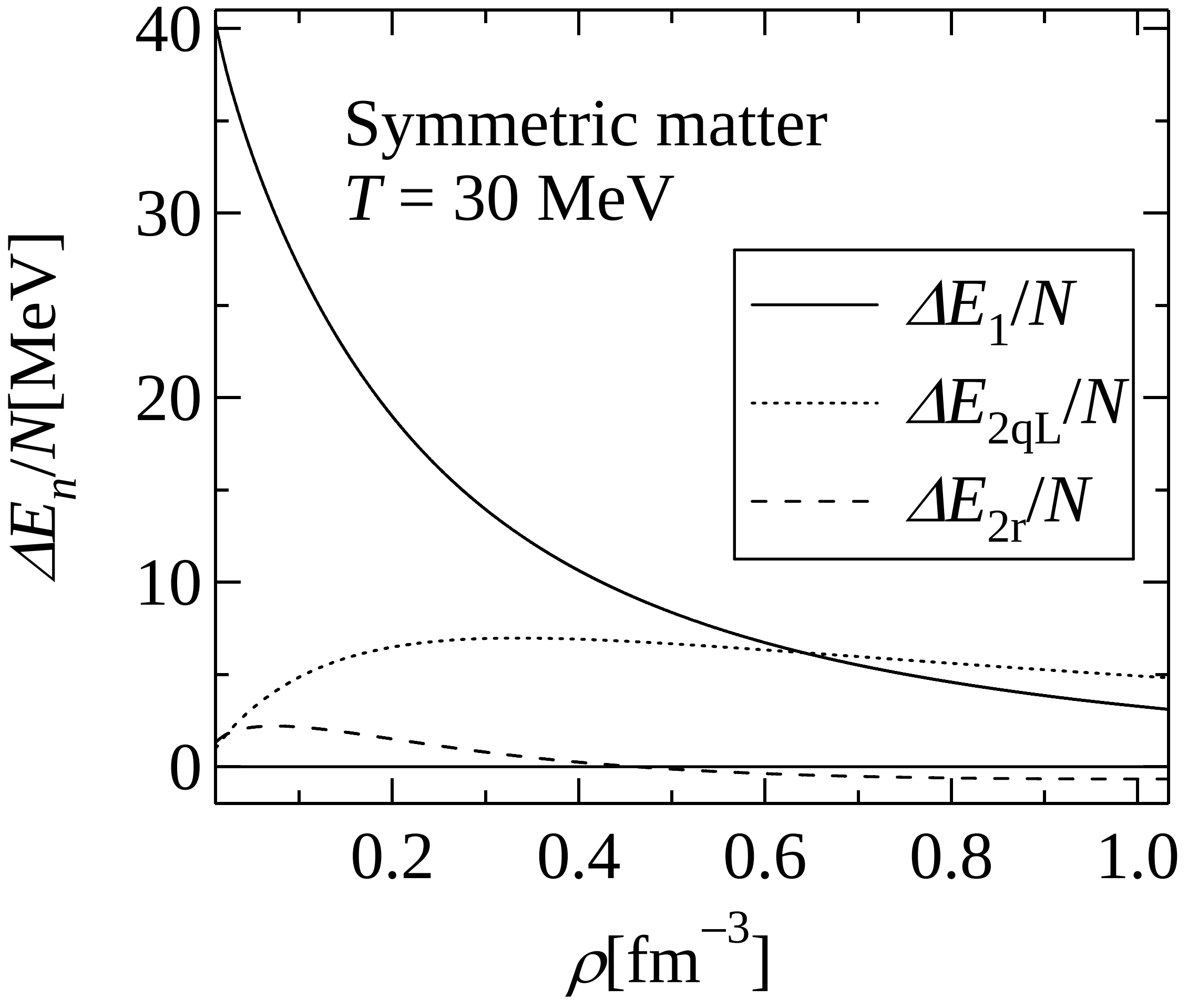

Fig.11